\newif\ifDRAFT \DRAFTfalse
\newif\ifPRE \PREfalse
\newif\ifXXX \XXXtrue

\ifDRAFT
 \documentclass[11pt]{article}
 \usepackage{fullpage}
 \usepackage[dvips]{graphicx}
 \usepackage{cite}
\fi

\ifPRE
 \documentstyle[twocolumn,aps,epsf]{revtex}
\fi

\ifXXX
 \documentclass[11pt]{article}
 \usepackage{fullpage}
 \usepackage[dvips]{graphicx}
 \usepackage{cite}
 \def\figwidth{3.5in}
\fi

\begin{document}


\def\bm#1{\mbox {\boldmath $ #1 $}}


\title{Phase Field Model for Three-Dimensional Dendritic Growth with
Fluid Flow}

\author{Jun-Ho~Jeong$^1$, Nigel~Goldenfeld$^2$ and Jonathan~A~Dantzig$^1$ \\
        $^1$Department of Mechanical and Industrial Engineering \\
        $^2$Department of Physics \\
	University of Illinois at Urbana-Champaign, Urbana, Illinois 61801}

\date{}

\maketitle

\begin{abstract}
We study the effect of fluid flow on three-dimensional (3D) dendrite
growth using a phase-field model on an adaptive finite element grid.
In order to simulate 3D fluid flow, we use an averaging method for
the flow problem coupled to the phase-field method and the
Semi-Implicit Approximated Projection Method (SIAPM).  We describe a
parallel implementation for the algorithm, using Charm++ FEM
framework, and demonstrate its efficiency.  We introduce an improved
method for extracting dendrite tip position and tip radius,
facilitating accurate comparison to theory.  We benchmark our results
for two-dimensional (2D) dendrite growth with solvability theory and
previous results, finding them to be in good agreement.  The physics
of dendritic growth with fluid flow in three dimensions is very
different from that in two dimensions, and we discuss the origin of
this behavior.
\end{abstract}

\section{Introduction}

Dendrites are the basic microstructural form for most crystalline
materials. They express the underlying crystalline symmetry, as
well as the growth conditions which existed when the dendrite was
formed. Dendrites may form from the vapor phase (e.g., snowflakes),
from solution (e.g., polymer crystals), or by solidification from the
melt (e.g., most metals). In this work, we focus our attention on
growth from the melt, which is important in many materials processing
applications.  Dendritic growth produces local compositional
variations which determine the macroscopic properties of the
material.  These features persist through subsequent processing, and
it is therefore important to understand the mechanisms by which the
microstructural pattern is selected.

Beginning with the morphological stability theory of Mullins and
Sekerka \cite{mullins}, the dynamics of pattern selection are now
reasonably well understood.  For a review of the theory, see Langer
et al. \cite{LanII80} and Kessler et al. \cite{Kes88} There is now a
consensus that the so-called ``microscopic solvability'' theory
\cite{Ben84,KessI84} agrees very well with numerical calculations.
\cite{Karma+Rappel,Provatas1998} These comparisons were performed in
two dimensions (2D), where accurate time-dependent simulations of
dendrite growth are tractable using phase-field methods, and also
level set techniques\cite{Tae} In this paper, we focus on the
phase-field method; for a description of the level set
method as applied to solidification problems, see Chen et
al.\cite{osher}

It is known that fluid flow during solidification dramatically alters
the solidification structure. \cite{davis} Using typical values for
the local flow velocity, material properties and process parameters,
one can anticipate that the interdendritic flow is dominated by
viscous forces (Re $\sim {\mathcal{O}}(0.1)$), but that the diffusion
fields for temperature and solute are dominated by advective effects
(Pe $\sim {\mathcal{O}}(10-100)$). The presence of the flow admits the
possibility of instabilities due to the flow itself, in addition to
the morphological instabilities normally found in crystal growth.

The effect of fluid flow on dendritic growth is the object of this
research. This is an inherently three-dimensional phenomenon, as can
be seen in the schematic drawing in Figure \ref{2d-3d}. A pair of
dendrites is shown growing into a flow which is nominally
perpendicular to their primary growth direction. The mechanism by
which the flow alters the growth pattern is the transport of solute
from the leading edge to the trailing edge of the dendrite.
\cite{Dantzig} In 2D, this occurs by the flow going up and over the
dendrite. However in three dimensions (3D), it is much easier for
this transport to take place by having the flow go \emph{around} the
dendrite. Thus, in order to correctly model this phenomenon, we must
do 3D simulations.

\ifXXX
\begin{figure}[htb]
 \centering
 \includegraphics[width=\figwidth]{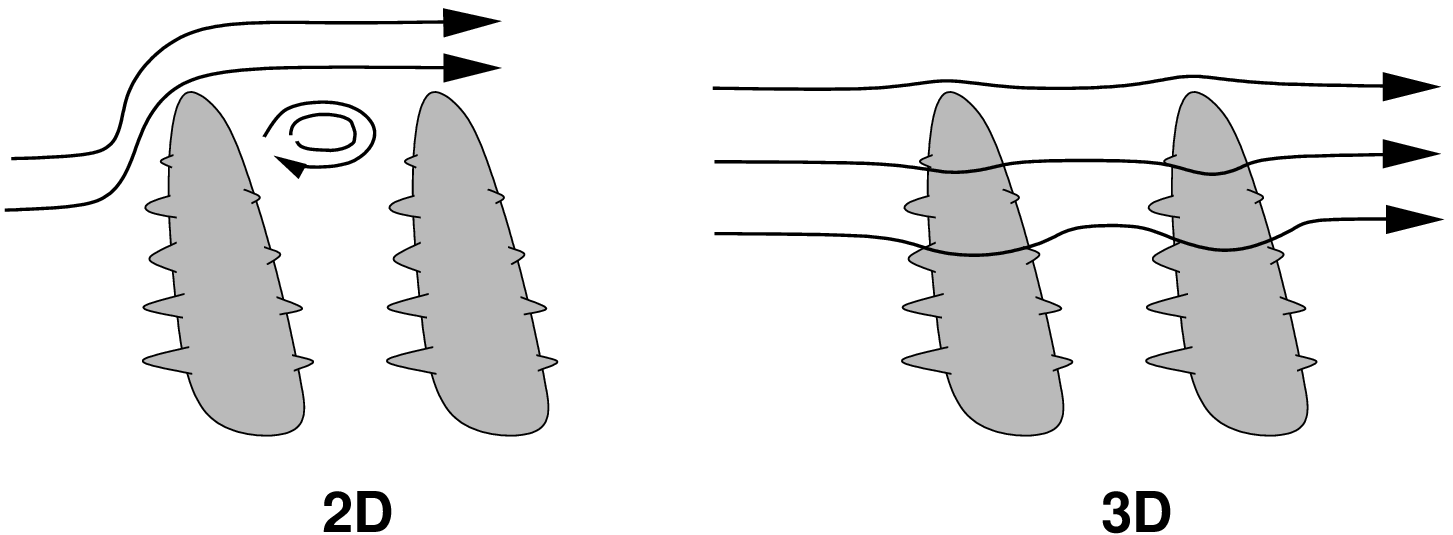}
\caption{Schematic drawing of the flow over dendrites growing
perpendicular to a superimposed flow, comparing 2D and 3D phenomena.}
\label{2d-3d}
\end{figure}
\fi


An example of a dendrite computed with fluid flow present is given
in Figure \ref{dendrite1}.  This calculation was done using the
methods we describe later in this paper, but we introduce it here
to provide a context for describing the physical problem. The
shape of the dendrite is complex, and it evolves during the
computation. The far-field flow on the left-hand side of the dendrite
is a uniform velocity, directed parallel to a preferred crystalline growth
direction.
The figure shows streamtraces for the flow over the
dendrite. Notice that growth is enhanced in the directions counter
to the flow. Sidebranches also appear preferentially on the leading
edge of the transverse arms, and the trailing arm is completely
suppressed. More will be said about this in Section \ref{results}.

\ifXXX
\begin{figure}[htb]
 \centering
 \includegraphics[width=\figwidth]{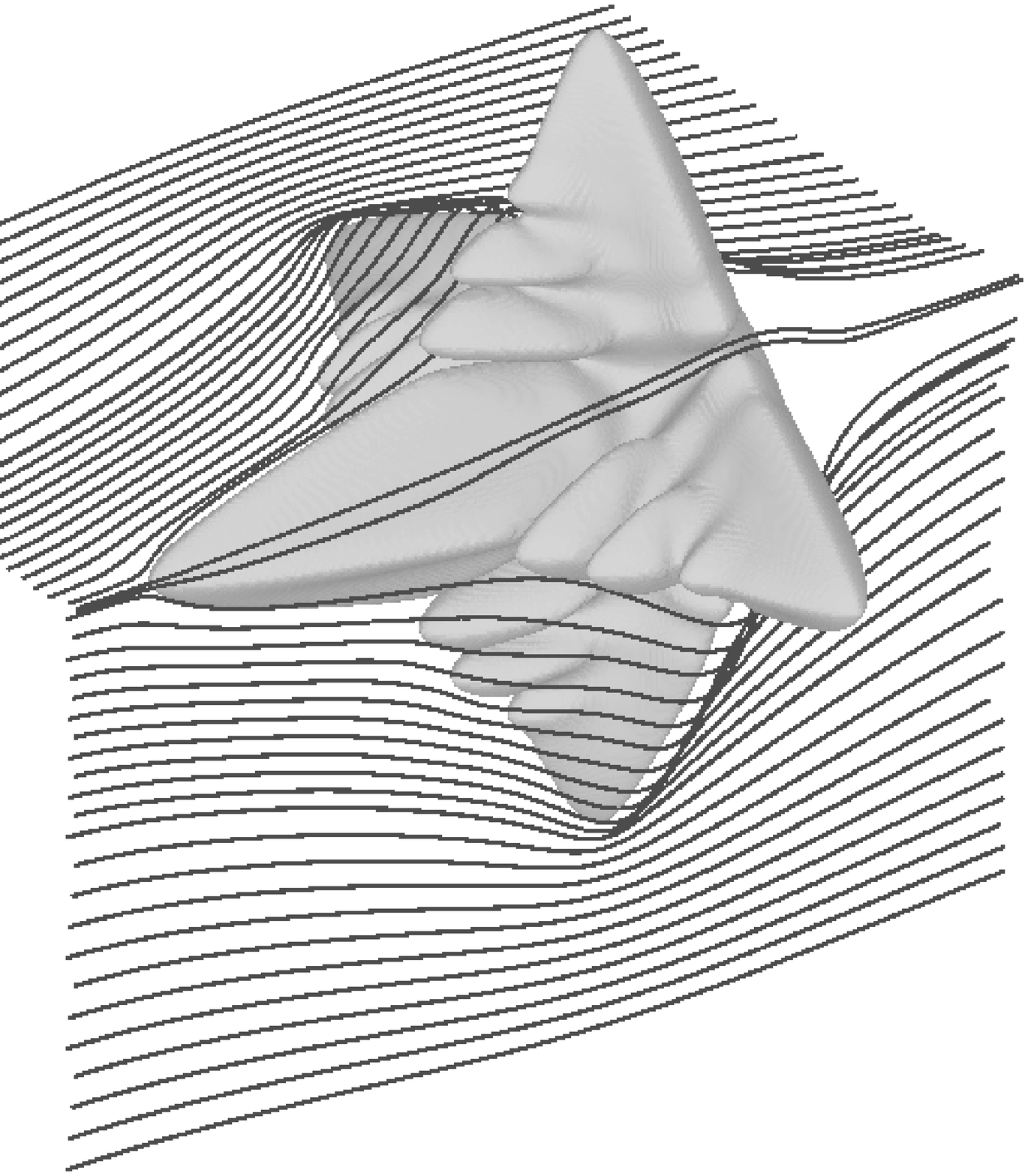}
\caption{Computed streamtraces for flow over a growing isolated
dendrite.}
\label{dendrite1}
\end{figure}
\fi

The surface of the dendrite represents the interface between the solid
and liquid, and there are two boundary conditions which must be
satisfied on this interface. First there is the condition of local
thermodynamic equilibrium
\begin{equation}
T = T_m - \Gamma({\bm n}) \kappa - \beta({\bm n}) {\bm V}\cdot{\bm n}
\end{equation}
where $T$ is the interface temperature, $T_m$ is the equilibrium
melting point of the pure material (with a flat interface), $\Gamma$
is the ratio of the surface energy to the entropy of fusion, $\kappa$
is the Gaussian curvature of the surface, $\beta$ is a kinetic
coefficient, ${\bm V}$ is the interface velocity and ${\bm n}$ is the
normal vector to the interface. The dependence of $\Gamma$ and
$\beta$ on ${\bm n}$ introduces crystalline anisotropy into the
problem. There is also an energy balance for the motion of the
interface
\begin{equation}
k_s \nabla T_s \cdot {\bm n} - k_\ell \nabla T_\ell \cdot {\bm n} = \rho_s L_f
{\bm V}\cdot{\bm n}
\end{equation}
where the subscripts $s$ and $\ell$ refer to the solid and liquid
phases respectively, $k$ is the thermal conductivity, $\rho_s$ is the
solid density and $L_f$ is the latent heat of fusion. We refer to the
solidification problem where these boundary conditions are explicitly
satisfied as the \emph{sharp interface problem}.

One of the computational issues in this problem is that the
position of the interface is \emph{a priori} unknown, and therefore
enforcement of the boundary conditions is difficult. Rather than track
the phase boundary explicitly, we introduce a continuous order
parameter $\phi \in [-1,1]$, where $\phi=-1$ corresponds to the
liquid, $\phi=1$ corresponds to the solid, and the level set $\phi=0$ is
identified as the interface. Details of the method, and the selection
of parameters to ensure convergence of the phase-field model to the
sharp interface problem described above, can be found in the references.
\cite{Karma+Rappel,Provatas1999}

The phase-field method introduces a finite width $W_0$ for the
interface, which must be kept small if the calculations are to be
meaningful. In particular, we require that $W_0$ must be of the order of
the capillary length $d_0$, which is a material property, typically
ranging from $1\times10^{-9}$ to $1\times10^{-8}$ m. At the same time,
the computations must resolve the diffusion field surrounding the
dendrite, and this can be of order $1\times10^{-4}$ m for physically
relevant growth conditions. Finally, the grid spacing at the interface
must be on the order of $W_0$ to preserve contact with the sharp
interface model. Thus, the spatial grid must resolve at least five
orders of magnitude. Uniform grid approaches are clearly limited to two
dimensions, and even then they are very computationally intensive.

In this paper, we resolve this difficulty by solving the phase-field 
equations on
an adaptive finite element mesh. The methods are discussed in greater
detail in the following sections; just a sketch is provided here. The
3D domain is meshed with hexahedral elements, stored in an octree data
structure. Local error estimators are used to selectively refine or
coarsen the mesh, and this permits tracking of the interface as well
as resolution of gradients in the other fields. There are six degrees
of freedom at each node (three velocities, pressure, temperature and
$\phi$), and a typical computation, such as the one shown in Figure
\ref{dendrite1} has up to 300,000 nodes, and thus well over one
million unknowns. While adaptive grid methods make the computations
feasible, the full 3D problem remains a formidable challenge.

The outline of this paper is as follows: In Section 2 we describe
numerical methods. In Section 3 we present a detailed description of
the 3D adaptive grid refinement algorithm. Section 4 describes a
parallel implementation using Charm++, and presents the results of
our effort to accelerate our code.  In section 5 we explain the
difficulties generically encountered in measuring interface
velocities, even with phase field methods, and present accurate
schemes to calculate the interface velocity.  In Section 6, we
present results for 3D dendrite growth without and with fluid
flow. In section 7, we conclude and discuss our results.

\section{Numerical methods}

The numerical implementation of the adaptive grid technique applied
to the phase-field model without convection has been described in
detail elsewhere, \cite{Provatas1999} so we focus here on the fluid
flow problem. Beckermann et al. \cite{Beckermann} introduced an
averaging method for the flow problem coupled to the phase-field, and
we follow that approach here. Beckermann et al. performed only 2D
calculations, but their methods extend naturally to 3D.  The phase
average $\Psi_k$ of a variable $\Psi$ for phase $k$ over volume
$\Delta$V, is defined as
\begin{equation} \Psi_k=\frac{2 \int_{\Delta V} X_k \Psi dV }{1-\phi}
\label{average}
\end{equation}
where $X_k \in [0,1]$ is an existence function. The phase averages of
the velocity and pressure are used for deriving the mixture
continuity equation, averaged liquid momentum equation, and averaged
energy conservation equation.  The formulation ensures that the fluid
velocity is extinguished in the solid, and further that the shear
stress at the liquid-solid interface is handled correctly.

The governing equations for simulating dendritic growth with fluid
flow are the mixture continuity equation, averaged liquid momentum
equation, averaged energy conservation equation, and phase field
equation as follows:

\begin{itemize}
\item
The mixture continuity equation:
\begin{equation} \nabla \cdot \left[ \frac{1-\phi}{2} {\bm u} \right]=0
\label{continuty}
\end{equation}
where ${\bm u}$ is the velocity vector.

\item
The averaged momentum equation:
\begin{eqnarray}
\frac{\partial}{\partial t} \left[ \left(\frac{1-\phi}{2}\right) {\bm u}
\right] + {\bm u} \cdot\nabla \left[ \left(\frac{1-\phi}{2}\right)
                 {\bm u} \right]+ \left(
\frac{1-\phi}{2} \right) \nabla p \nonumber \\
= \nu \nabla^2
\left[\left(\frac{1-\phi}{2} \right) {\bm u} \right] -\nu \frac{h^2
(1-\phi^2)(1+\phi)}{8 \delta^2}{\bm u}
\label{momentum}
\end{eqnarray}
where $t$ is time, $p$ is pressure, $\nu$ is the kinematic viscosity,
$\delta = W_0/\sqrt{2}$ is the characteristic interface width, and $h$ is a
constant (=2.757) which ensures that the interface shear stress is
correct for a simple shear flow (see Beckermann et al.
\cite{Beckermann}).

\item
We write the averaged energy conservation equation in terms of a
dimensionless temperature $\theta = c_p(T-T_m)/L_f$, scaled by the
specific heat $c_p$ and latent heat of fusion $L_f$:
\begin{equation}
\frac{\partial \theta}{\partial t} + \left(\frac{1-\phi}{2}\right){\bm u}
\cdot\nabla \theta = D \nabla^2 \theta + \frac{1}{2} \frac{\partial 
\phi}{\partial t}
\label{energe}
\end{equation}
where $D=\alpha \tau_0 / W_0^2$ in which $\alpha$ is the thermal
diffusivity and $\tau_0$ is a time characterizing atomic motion in
the interface.

\item
The 3D phase-field evolution equation is given by
\begin{eqnarray}
\tau ({\bm n}) \frac{\partial \phi}{\partial t} & = &
   \left[\phi-\lambda \theta (1-{\phi}^2) \right](1-\phi^2)+\nabla \cdot
   \left[W({\bf n})^2 \nabla \phi \right] \nonumber \\ 
& + & \partial_x \left(
   \left| \nabla \phi \right|^2 W({\bf n}) \frac{\partial W({\bf
   n})}{\partial (\partial_x \phi)}   \right) \nonumber \\ 
& + & \partial_y \left(
   \left| \nabla \phi \right|^2 W({\bf n}) \frac{\partial W({\bf
   n})}{\partial (\partial_y \phi)}   \right) \label{phase} \\ 
& + & \partial_z
   \left( \left| \nabla \phi \right|^2 W({\bf n}) \frac{\partial W({\bm
   n})}{\partial (\partial_z \phi)}   \right) \nonumber 
\end{eqnarray}
where $\lambda$ is a dimensionless constant that controls the tilt of
the double well potential which forces $\phi$ to the attractors at $\pm
1$. Anisotropy is included in this equation by writing the interface
mobility $\tau$ and width $W$ as functions of the local normal vector
${\bm n}$. Following Karma and Rappel \cite{Karma+Rappel}, we choose
\begin{equation}
 W({\bm n})=W_0 a_s({\bm n}) ; \qquad
 \tau({\bm n})=\tau_0 a_s^2({\bm n})
\end{equation}
with
\begin{equation}
 a_s({\bm n})=(1-3\epsilon_4)
 \left[1+\frac{4\epsilon_4}{1-3\epsilon_4} \frac{(\partial_x
 \phi)^4+(\partial_y \phi)^4+(\partial_z \phi)^4}{\left| \nabla
 \phi \right|^4} \right]
\end{equation}
The constant parameter $\epsilon_4$ fixes the strength of
the anisotropy in the interface energy.
\end{itemize}

We solve the 3D flow equations using the Semi-Implicit Approximate
Projection Method (SIAPM) as developed by Gresho \cite{Gresho1995}. SIAPM
is a predictor-corrector method which can solve Eq.~(\ref{continuty})
and Eq.~(\ref{momentum}) effectively, especially for large 3D problems,
because it uses relatively small amounts of memory. The velocity degrees
of freedom are solved in a segregated form, and the pressure is updated
using a projection method. For a detailed discussion of the algorithm,
the reader is referred to the original paper \cite{Gresho1995}, and we
present only an operational description of the algorithm here. The
algorithm consists of four steps:

\begin{enumerate}
\item
Compute an intermediate velocity ${\tilde{\bm u}^{n+1}}$ from
\begin{eqnarray}
\left( \frac{1}{\Delta t}{\bm M}-\frac{1}{2}{\bm K} + {\bm F}
\right) \tilde{{\bm u}}_i^{n+1}=
		 \left( \frac{1}{\Delta t}{\bm M}+\frac{1}{2}{\bm K}
		 \right) {\bm u}_i^n \nonumber \\
-{\bm A}({\bm u^n}){\bm u}_i^n-{\bm G}_i {\bm p}^n 
\label{intermediate} 
\end{eqnarray}
where $\tilde{{\bm u}}_i^{n+1}$ is the vector of nodal values of
the intermediate
velocity component $i$ at timestep $n+1$, ${\bm u}_i^{n}$
is the corresponding vector at timestep $n$, and ${\bm p}^n$ is the
vector of nodal pressures at timestep $n$.
The coefficient matrices
are defined in terms of the velocity shape functions ${\bm N}$ 
as follows:
\begin{eqnarray}
&{\bm M} & = \int_\Omega \frac{(1-\phi)}{2} {\bm N}^T{\bm N} d\Omega \\
&{\bm K} & = \int_\Omega \nu
\frac{(\phi-1)}{2} \left( \frac{\partial {\bm N}^T}{\partial {\bm x}}
\frac{\partial {\bm N}}{\partial {\bm x}} + {\bm I} \frac{\partial
{\bm N}^T}{\partial x_k} \frac{\partial {\bm N}}{\partial x_k} \right)
d\Omega \\
& {\bm F} &= \int_\Omega {\nu h (1-\phi^2)(1+\phi) \over 8 \delta^2}
{\bm N}^T {\bm N} d\Omega \\
&{\bm A}({\bm u}^n)& = \int_\Omega 
\frac{(1-\phi)}{2} {\bm N}^T u_k^n \frac{\partial {\bm N}}{\partial
  x_k} d\Omega \\
& {\bm G}_{i} &= \int_\Omega
\frac{(1-\phi)}{2} {\bm N}^T \frac{\partial { \bm N}}{\partial x_i}
d\Omega
\end{eqnarray}

\item
The velocity field found in the first step is generally not divergence-free.
The next step corrects the pressure to obtain an approximately
divergence-free velocity field by solving a Poisson equation for 
${\Delta {\bm p}^{n+1}=\left({\bm p}^{n+1}-{\bm
p}^{n}\right)}$:
\begin{equation}
{\bm L} \Delta {\bm p}^{n+1} = -{1 \over \Delta t}{\bm D}\left( \tilde{{\bm
u}}^{n+1}-{\bm u}^n \right)
\label{pressure}
\end{equation}
where
\begin{eqnarray}
{\bm L}&=&\int_\Omega \frac{(1-\phi)}{2}
\frac{\partial {\bm N}^T}{\partial x_k} \frac{\partial
{\bm N}}{\partial x_k} d\Omega \\
{\bm D}&=&\int_\Omega \frac{(1-\phi)}{2} {\bm N}^T \frac{\partial
{\bm N}}{\partial {\bm x}} d\Omega 
\end{eqnarray}
${\bm p}^{n+1}$ is then updated from
\begin{equation}
{\bm p}^{n+1} = {\bm p}^n + \Delta {\bm p}^{n+1}
\end{equation}

\item
Finally, the projected velocity ${{\bm u}^{n+1}}$ is computed in a
corrector step by solving
\begin{equation}
{\bm u}^{n+1} = \tilde{{\bm u}}^{n+1}-\Delta t {\bm
M_L}^{-1} {\bm G}\Delta {\bm p}^{n+1}
\end{equation}
where
\begin{equation}
{\bm M_L} = \int_\Omega {1-\phi \over 2} {\bm N} d\Omega
\end{equation}
\end{enumerate} 

The computations are started using an initial velocity field ${{\bm
u}_0}$
determined from the boundary conditions.
In order to obtain a field ${{\bm u}_0}$ whose discrete divergence is close
to zero, we perform {\tt J} iterations (typically 10) iterations on the
following system (shown in pseudo-code):
\begin{eqnarray}
 \tt {For~I}~=~{1~to~J~[} \nonumber \\
 \tt Lq^{I} &=& \tt -Du^{I-1} \\ 
 \tt u^{I}  &=& \tt u^{I-1}-M_{L}^{-1}Gq^{I} \nonumber
\hspace{2em} ]
\end{eqnarray} 
The variable $\tt{q}$ plays the role of a temporary pressure update,
but the actual initial pressure is zero everywhere.

Eqs.~(\ref{intermediate}) and (\ref{pressure}) are
solved by the conjugate gradient (CG) method with diagonal
preconditioning\cite{Mikic}.  The SIAPM can calculate the
velocity field for large 3D problems much faster than fully
implicit time-stepping methods, because convergence is
reached for Eq.~(\ref{intermediate}) in a few iterations and
the number of degree of freedom of Eq.~(\ref{pressure}) is one.
The CG iteration for Eq.~(\ref{pressure}) converges more slowly,
typically 50-200 iterations.

The averaged energy equation Eq.~(\ref{energe}) is also solved using
the CG method with diagonal preconditioning. Streamline upwind
schemes \cite{Brooks} are employed for the convection terms in
Eq.~(\ref{intermediate}) and Eq.~(\ref{energe}).  The 3D phase field
equation is a nonlinear system. In order to solve the system
implicitly, an iterative method such as the Newton-Raphson method is
required.  We use instead an explicit time-stepping scheme where a
linear system is solved.  Stable solutions are obtained from the
explicit scheme, because the variation of the $\phi$ field exists
only in the interface region and a sufficiently small time increment
$\Delta$t is used.

\section{3D adaptive grid refinement algorithm}

To resolve the interface, the grid spacing $\Delta$x must be
smaller than the characteristic interface width $W_0$, which must in turn
be on the order of the capillary length, for the solution of the
phase field model to converge to the sharp interface limit.  
On the other hand, the system size $L$ required for simulation is
determined by the size of the diffusion field ahead of the dendrite.
A ratio of $L/{\Delta x \sim 10^3-10^4}$ is typical.
For simulation of a single 3D dendrite,
this implies that at least 10 million elements must be used in a uniform
grid.  Such a simulation of 3D dendrite growth with fluid flow
would be intractable.

In this problem, there is an important characteristic that the
various fields vary most rapidly in the interface region, whose
width is much smaller than $L$.  For this reason,
adaptive grid refinement techniques can be applied very effectively.
The 3D adaptive grid refinement is described in the next 
few sections.

\subsection{Error estimating procedure}

The basis of the code is the element data structure, illustrated in
Figure \ref{link}.  The structure consists of arrays for element
connectivity, neighbors and also the element pressure.

\ifXXX
\begin{figure}[htb]
 \centering
\scriptsize{
\begin{verbatim}
type element
  integer                         :: num_element            !! Element number
  integer                         :: level                  !! Refinement level
  integer                         :: lneigh                 !! Number of neighbor elements
  type(connectivity), pointer     :: connect                !! Pointer of connectivity		
  type(connectivity), pointer     :: connect_mid            !! Pointer of connectivity for disconnected nodes
  integer                         :: midindex(6)            !! Index to check for discontinuous nodes
  type(neighbor_elements), pointer:: neighbor               !! Pointer to neighbor elements
  integer                         :: num_parent(LimitLevel) !! Parent element numbers
  integer                         :: num_history(LimitLevel)!! Time step number 
  integer                         :: merge                  !! Index to check if the element should be merged
  real*8                          :: error                  !! Error estimator value at time step n
  integer                         :: nver                   !! Vertex node number for disconnected edge node
  integer                         :: ntype                  !! Element type (liquid/solid/interface)
                                                            !! If phi = -1         , ntype = 1(liquid)
                                                            !! If phi = 1          , ntype = 2(solid)
                                                            !! If -1 < phi < 1     , ntype = 3(interface)
  real*8                          :: pe                     !! Element pressure
  type(element), pointer          :: previous               !! Previous element in linked list
  type(element), pointer          :: next                   !! Next element in linked list
end type element
\end{verbatim}
}
\caption{Element linked data structure for adaptive grid}
\label{link}
\end{figure}
\fi

The grid is locally adapted based on an element-by-element error
estimate, with a hybrid scheme using the magnitude of $\phi$ and the
interelement variation of the derivatives of $\theta$.  We use $\phi$
as an indicator to define the specific region in which the finest
elements should be distributed.  Specially, if an element includes a
node where
\begin{equation}
\phi_{min} \le \phi \le \phi_{max}
\end{equation}
then the element is divided until its refinement level becomes the
maximum level.  We control the width of the region with the finest
elements through the values of $\phi_{min}$ and $\phi_{max}$.  We
proceed by defining a grid, and then solving a predetermined number
of time steps on that grid $N_{ref}$ (typically $N_{ref}$=20-100),
and then adapt the grid. We require that the interface remains within
the fine grid during the time steps. We found that $\phi_{min} =
-0.99$ and $\phi_{max}=0.9$ gave consistent results.  The reason for
the asymmetry is that the interface moves from the solid region $\phi
> 0$ to the liquid region $\phi < 0$.  Outside of the interface
region ($\phi < \phi_{min}$ or $\phi > \phi_{max}$), we used an error
estimator based on the magnitudes of the derivatives of $\theta$ as
follows:

\begin{equation}
E_e = \int_{{\Omega}_e} \nabla \theta \cdot \nabla \theta d \Omega
\end{equation}
where ${\Omega}_e$ is the element.
The estimated error ${E_{sub}}$ of a subelement divided from a parent
element with error $E_p$ can be calculated from the asymptotic rate of
convergence of the finite element approximation as 

\begin{equation}
E_{sub} = {\left( \frac{h_{sub}}{h_{parent}} \right) }^k E_p = 0.5^k E_p
\end{equation}
where $k$ is the exponent for the asymptotic rate of convergence ($k=2$ for 
linear elements), and $h_{sub}$ and ${h_{parent}}$ are the element sizes
of the subelement and parent element, respectively.
A limit value ${E_{limit}}$ is calculated from
\begin{equation}
E_{limit} = \gamma \max_{e=1}^{N_T} \left[ \left(0.5^k
\right)^{\left(N_{max}-N_{e}\right)} E_p \right]
\end{equation}
where ${\gamma}$ is a scale coefficient (typically 10), $N_T$ is
the total number of elements, $N_{max}$ is maximum refinement level, and
$N_e$ is the refinement level of element $e$.
The element is divided until its estimated error becomes smaller than 
a specified limit value. Once the elements to be refined have been selected,
the grid is refined using the procedure described in the next section. 

\subsection{Grid refinement procedure}

The grid refinement procedure begins by storing the pointers of
neighboring elements and the refinement level of each element.
Refinement is required in an element whenever the estimated error
exceeds the limit value, or when the absolute difference between the
level of an element and that of its neighbors exceeds one. The latter
is called the single-level rule.  Refinement is carried out
successively at each refinement level, beginning with the minimum,
according to the procedure illustrated in Figure \ref{refine}, and
described below.

\ifXXX
\begin{figure}[htb]
 \centering
 \includegraphics[height=\figwidth,angle=0]{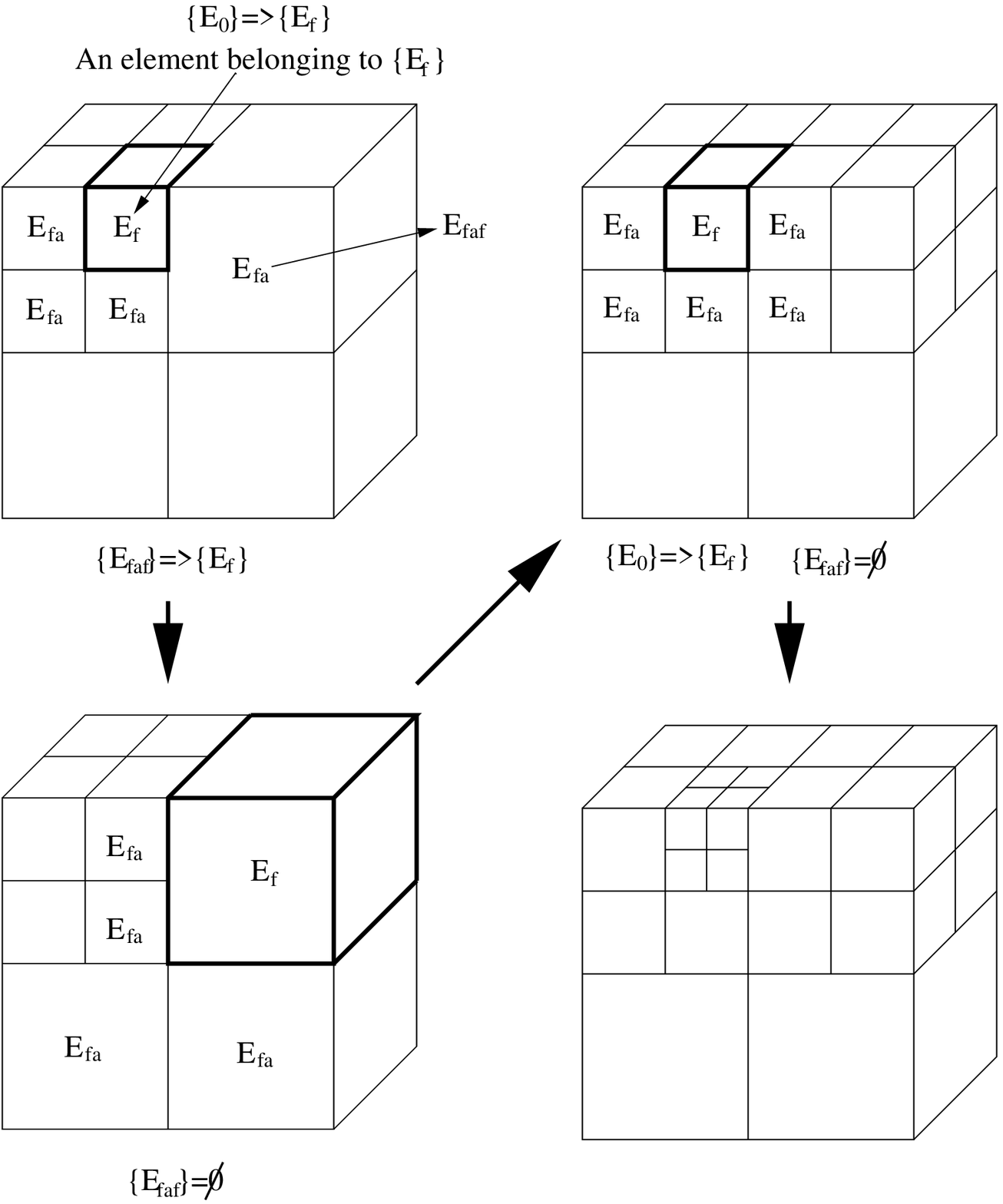}
\caption{Dividing sequence for refinement. The arrows indicate the
sequence of refinement, as discussed in the text.}
\label{refine}
\end{figure}
\fi

\begin{itemize}
\item
After checking the computed elemental errors, a set
${\{E_0\}}$ is created which consists of elements in the current
refinement level, whose error exceeds the present limit value. These
elements will be refined after the neighboring elements which must be
refined to satisfy the single-level rule are found and refined.

\item
A recursive search is then performed to find the outermost elements
which need to be refined to satisfy the single-level rule. We do this
by first defining a new set ${\{E_f\}}$ whose initial value is
${\{E_0\}}$. The neighbor elements to each element in $\{E_f\}$ are
then examined, and any neighboring element that does not satisfy the
single-level rule is added to a new set ${\{E_{faf}\}}$.  If
${\{E_{faf}\}}$ is not a null set then ${\{E_f\}}$ is replaced by
${\{E_{faf}\}}$. This procedure is repeated until ${\{E_{faf}\}}$
becomes a null set. At this point, the set $\{E_f\}$ contains the
outermost elements which need to be refined to satisfy the
single-level rule, and these elements are then divided into eight
subelements by bisection of each face. (See Figure \ref{refine}.) At
this point, the set $\{E_f\}$ is again set equal to $\{E_0\}$, and
the search begins anew.  This process is repeated until the search
for elements to fill set $\{E_{faf}\}$ yields a null set. Then, the
elements in set $\{E_0\}$ are refined, and the refinement process is
completed in the next step.

\item
The nodal coordinates, element connectivities, neighbor arrays,
parent arrays, nodal refinement levels, etc.~are updated. The
neighbor array is the array to store the element numbers of
neighboring elements. The parent array is the array to store the
element number of the parent elements for the coarsening procedure,
described in the next section.

\end{itemize}

The recursive search for elements violating the single-level rule in
the second step above may seem inefficient, because it performs
multiple searches, finding the same elements. However, this procedure
makes it possible to perform the searches with each element knowing
only about its nearest neighbors. This gives greater efficiency in the
computational phase, because it limits the memory allocation required
within the element data structure.

\subsection{Grid unrefinement procedure}

The unrefinement procedure is accomplished through a loop in which the 
refinement level is decreased incrementally from the maximum level to the
minimum level.

\begin{itemize}
\item
A set ${\{\{E_{sf}\}\}}$ is created, containing all elements which
are eligible for unrefinement based on the value of the error
estimator.  Each element of ${\{\{E_{sf}\}\}}$ is a subset
${\{E_{sf}\}}$, consisting of eight elements to be merged into one.
If an element whose refinement level equals the current level of the
unrefinement loop has smaller estimated error than a prescribed limit
value, then the element is placed in a temporary set
${\{E_{temp}\}}$.  The elements belongs to ${\{E_{temp}\}}$ are
sorted into the subsets ${\{E_{tsf}\}}$ according to their parent
element.

\item
If the number of elements belonging to any subset
${\{E_{tsf}\}}$ is eight and an element created from the eight
elements satisfies the single level rule, then ${\{E_{tsf}\}}$
becomes the subset ${\{E_{sf}\}}$.
\item
The eight elements belonging to each ${\{E_{sf}\}}$
are then merged into the parent element. As we did for grid refinement, the
nodal coordinates, elemental connectivities, neighbor array, parent
array, and nodal refinement levels are updated.
\end{itemize}

\subsection{Treatment of disconnected nodes}

The adapted grid will have so-called disconnected nodes which appear
whenever an element has a neighbor whose refinement level differ by
one, as illustrated in Figure \ref{discon}.  An element with
connectivity [6, 3, 12, 4, 1, 2, 13, 5] contacts two neighbor elements of
lower refinement level.

\ifXXX
\begin{figure}[htb]
 \centering
 \includegraphics[width=\figwidth]{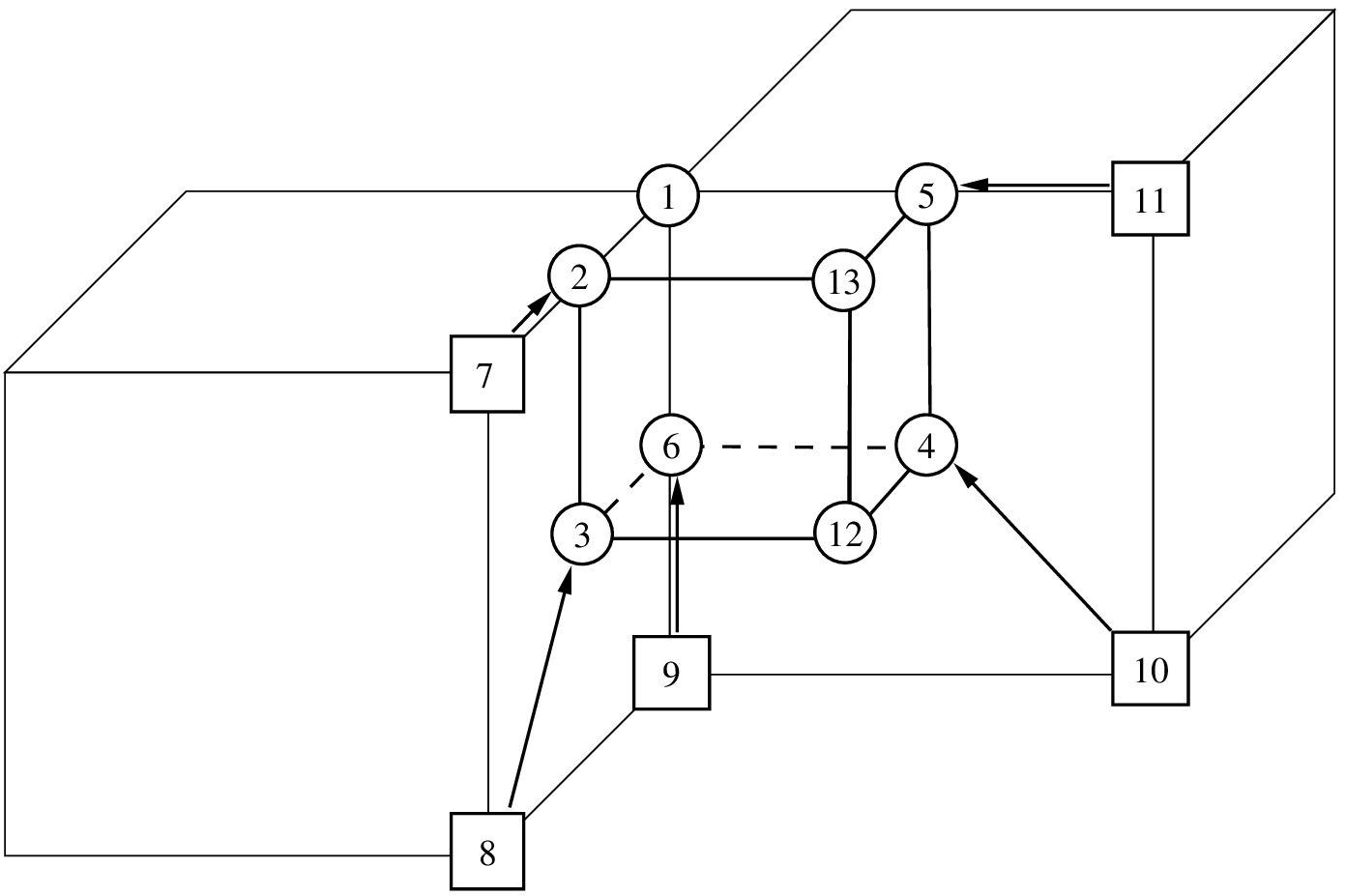}
\caption{Configurations of disconnected sides and nodes}
\label{discon}
\end{figure}
\fi

To ensure continuity of the solution between elements, the following constraints
are enforced for each degree of freedom, here represented by the symbol $v_i$,
\begin{itemize}
\item
Disconnected edge mid-nodes:
\begin{equation}
v_2 = \frac{v_1+v_7}{2}, \hspace{0.3cm} v_5 = \frac{v_1+v_{11}}{2},
\hspace{0.3cm} v_6 = \frac{v_1+v_9}{2}
\label{edgemid}
\end{equation}
\item
Disconnected face mid-nodes:
\begin{equation}
v_3 = \frac{v_1+v_7+v_8+v_9}{4}, \hspace{0.1cm} v_4 =
\frac{v_1+v_{11}+v_{10}+v_9}{4}
\label{facemid}
\end{equation}
\end{itemize}
After applying the constraints, the original element connectivity is modified
to be [8, 12, 10, 9, 7, 13, 11, 1], and the element shape functions are changed
appropriately. This change is made to facilitate domain decomposition in the 
parallel implementation, described next.

\vspace{0.5cm}
\section{Parallelization by Charm++}

Charm++ is a message-passing parallel runtime system for machines
from clusters of workstations to tightly-coupled symmetric
multi-processing machines. \cite{charm} A parallel FEM code can be
written in FORTRAN90 using interface routines from the Charm++ FEM
framework. \cite{fem-hipc} The FEM framework program consists of
three kinds of subroutines.  Service routines such as $``$INIT" and
$``$FINALIZE" do I/O, startup and shutdown tasks, and are called only
on the first processor.  The main work is done using $``$DRIVER"
routines, replicated on every processor.  In our parallel
implementation, the adapted grid is newly regenerated every 20-100
time steps. After each regeneration, the adapted grid is partitioned
into chunks assigned to each processor using METIS. \cite{metis} This
function is also handled through the interface to Charm++.

In the DRIVER routine on each processor, the temperature and velocity
fields are calculated using the preconditioned CG method in an the
element-by-element scheme, and the $\phi$ field is solved using an
explicit time stepping scheme, as described earlier.  All
calculations for the CG method are accomplished through the products
of the elemental stiffness matrixes and local solution vectors.  This
creates additive contributions for the residual vector for each
degree of freedom. Some $``$shared nodes" appear on more than one
processor.  The array is calculated for each chunk, and the values of
the field variables for all shared nodes are combined with the other
chunks by using the Charm++ function $ \rm ``FEM\_Update\_Field$".
For calculating the inner product of arrays and finding the maximum
error values, each nodal value is combined across all chunks and the
shared nodes ares not double-counted by using the Charm++ function $
\rm ``FEM\_Reduce\_Field$".  After the computations in the DRIVER
routine on all processors are completed, the calculated data are
transfered from DRIVER to INIT.  The grid is adapted, then
repartitioned.  These procedures are repeated until the simulation is
finished.

In order to rate the parallel performance of the code,
we compute the ratio SP, defined as
\begin{equation}
\rm SP=\frac{run\;time\;on\;one\;processor}{run\;time\;on\;n\;processors}
\end{equation}
An example result is shown in Figure \ref{speedup}, where we used a grid with 
296,636 elements and 349,704 nodes computing over 20 time steps. 
The value SP=28.8 for 32 processors is typical for our code, and shows that
the code has been effectively parallelized.

\ifXXX
\begin{figure}[htb]
 \centering
 \includegraphics[width=\figwidth]{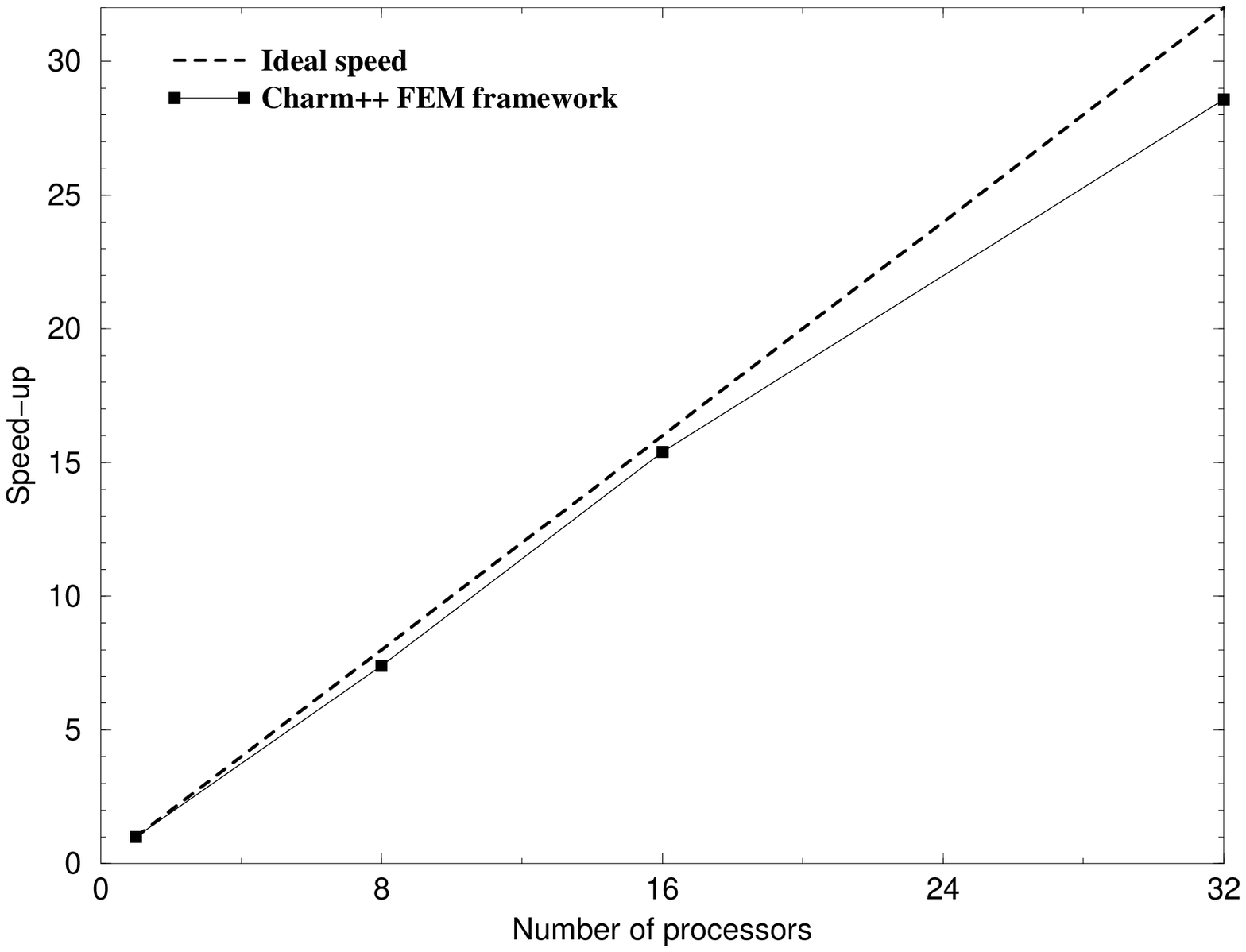}
\caption{Speed-up for Charm++ FEM framework}
\label{speedup}
\end{figure}
\fi

\section{Tip Velocity Measurement}

The steady state tip velocity is a convenient measure 
to compare the computational results to dendritic growth theories.
In the phase-field method, tip velocity
is inferred from the temporal evolution of $\phi=0$ along the primary
growth axis. Karma and Rappel \cite{Karma+Rappel} were able to derive
interface velocities from the numerical solution by using a third-order 
polynomial to interpolate $\phi$, using neighbor 
and next-neighbor grid point values. Using a 1-D analog problem as a
basis, they showed that this scheme gives an interface velocity with a small
amplitude oscillation about the mean, due to the interpolation of the moving
front on the fixed grid.

This multipoint scheme is somewhat problematic when adaptive gridding is used,
because the selection of points for interpolation is incompatible with
our data structure. We developed a new scheme using a hyperbolic tangent
function to perform the interpolation. This approach provides equivalent
control of the tip position oscillation, yet it requires only two grid
points, corresponding to the nodes in the element which contains the
interface ($\phi=0$). Let us denote the $x$-coordinates and $\phi$ values
of these two points as $(x_1,\phi_1)$ and $(x_2,\phi_2)$. We interpolate
$\phi$ along the axis as
\begin{equation}
\phi(x;x_c,W) = -{\rm tanh} \left( {x-x_c \over W} \right)
\end{equation}
where $x_c$ and $W$ are fitted parameters corresponding to the
zero-crossing of $\phi$ and the interface width, respectively. The
parameters $x_c$ and $W$ are determined as follows:
\begin{eqnarray}
 W  & = & {2(x_1-x_2) \over \ln[(1-\phi_1)(1+\phi_2)] -
           \ln[(1+\phi_1)(1-\phi_2)]}  \\
 x_c & = & {{W \over 2} \ln{1+\phi_1 \over 1-\phi_1}+x_1}
\end{eqnarray}
The interface velocity is then computed by a finite difference in time
between successive interface locations.

To demonstrate the scheme, we examined the same 1D test problem
considered by Karma and Rappel \cite{Karma+Rappel}, viz.
\begin{equation}
\tau_0 {\partial \psi \over \partial t} = W_0^2 
   {\partial^2 \psi \over \partial x^2} + \psi - \psi^3 + \Delta
\label{Front}
\end{equation}
using $\tau_0$=1, $W_0$=1, $\Delta$x=0.8 and $\Delta$=0.02.
This problem has a traveling wave solution which propagates at velocity
V=0.041. 
The computed interface velocity is shown in Figure \ref{TIPVEL1D} 
for a variety of interpolation schemes. It can be seen that the 
hyperbolic tangent interpolation scheme has a small oscillation amplitude, 
slightly smaller than that of the third-order polynomial. 
The average interface velocity can be easily extracted by selecting
pairs of interpolated points separated by the amount of time it takes to
cross one grid spacing. We refer to this as the moving average tip velocity.
This method is used to compute the interface velocities reported for the
calculations in the remainder of this paper.

\ifXXX
\begin{figure}[htb]
 \centering
 \includegraphics[width=\figwidth]{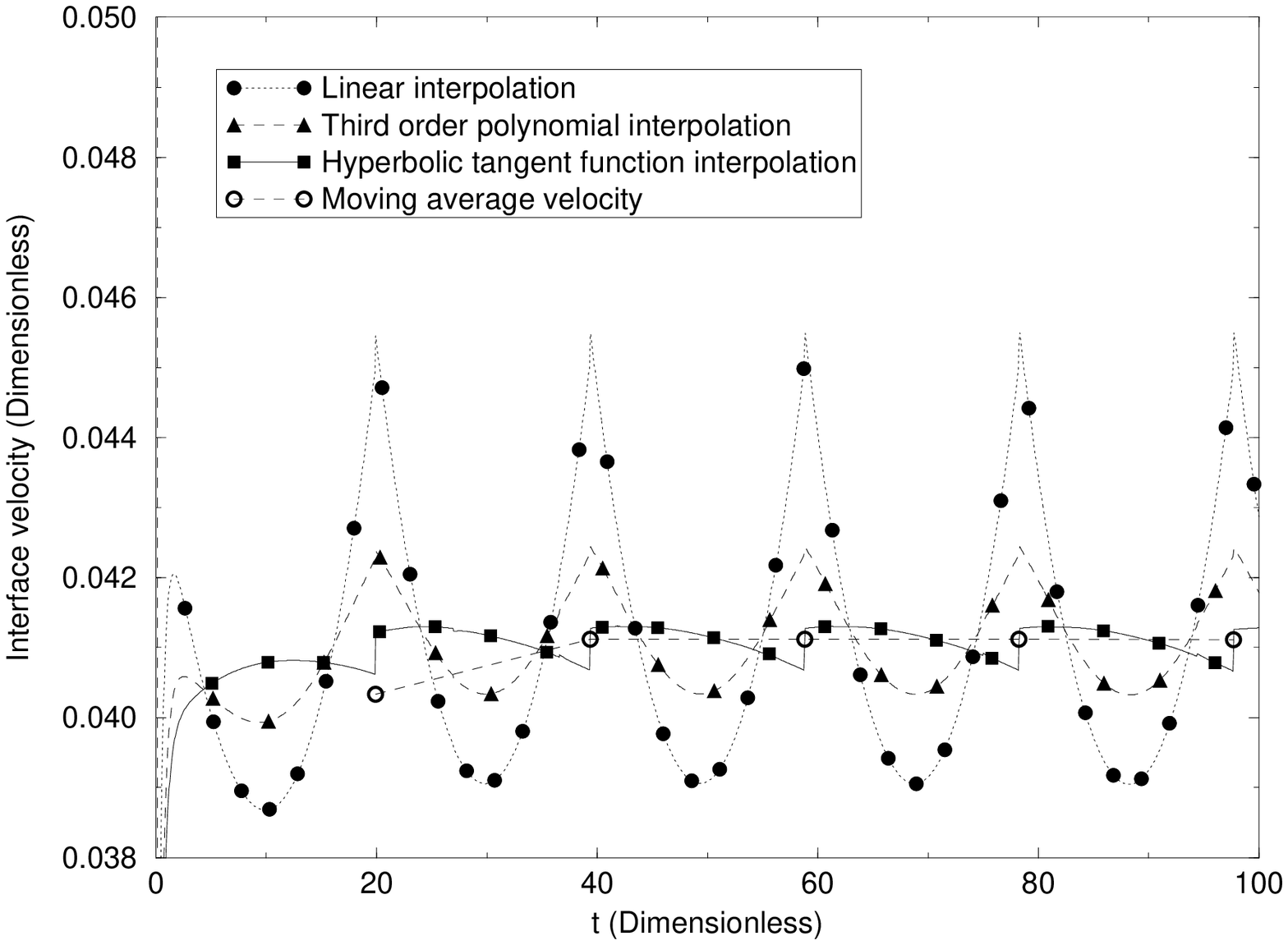}
\caption{Interface velocity versus time for 1-D test problem}
\label{TIPVEL1D}
\end{figure}
\fi

\section{Results}
\label{results}

\subsection{2D Verification Problem}

In order to validate our 3D code, we simulated the 2D example
analyzed by Beckermann et al.\cite{Beckermann}.  The computational
domain is a square with an initial circular seed.  A 3D domain is
created by extending the square domain and the circle seed into the
y-direction.

We used a box of edge length $L=204.8$ so that subelements of $\Delta
x=0.8$ or 0.4 are created through repeated bisection of the domain.
The flow enters at the left edge, with $u_x=1$ and $u_y=0$. The top
and bottom surfaces are symmetry boundaries. Beckermann et
al.\cite{Beckermann} used $L=230.4$, and similar values for
$\Delta$x.  The problem parameters are: undercooling $\Delta$=0.55,
thermal diffusivity D=4, and anisotropy $\epsilon_4$=0.05. We also
used $\Delta$t=0.016, $W$=1, $\tau_0$=1, and $\lambda$=6.383. The
capillary length $d_0$(=$a_1\frac{W}{\lambda}$) is 0.139, the
kinematic viscosity $\nu$ is 92.4 and the Prandtl number Pr
calculated from these parameters is 23.1. These physical parameters
correspond to succinonitrile.  The adapted grid is newly regenerated
every 20 time steps. We examined two minimum grid spacings, $\Delta
x_{\rm min}$=0.8 and $\Delta x_{\rm min}$=0.4.  The largest element
size was $\Delta x_{\rm max}=3.2$ for both cases.

For 2D dendrite growth without fluid flow, when using $\Delta x$=0.8
and $\Delta x$=0.4, the steady state dendrite tip velocities $V_{tip}$
scaled with $D/ d_0$ are 0.01744 and 0.01689, respectively. Those
values are in good agreement with solvability theory (0.01700) and
previous phase-field results.\cite{Beckermann}  We calculated the tip
radius $\rho_{tip}$ using the method in Reference \cite{Karma+Rappel},
where the tip curvature is computed using estimates of the second
derivatives at the tip interpolated along the two coordinate
directions at the tip.  We obtained $\rho_{tip} / d_0$=10.58 and 8.78
for $\Delta x$=0.8 and $\Delta x$=0.4, respectively. The ratio
$\rho_{tip} / d_0$ for $\Delta x \rightarrow 0$ can be computed by
Richardson extrapolation, plotting $\rho_{tip}$ versus ${\Delta
x}^2$. The extrapolated value of this ratio for $\Delta x \rightarrow
0$ is 8.20, which is somewhat larger than the solvability solution
(6.90).

An alternative method to fit the tip radius, given by Wheeler
\cite{Wheeler}, was also used. In this method, we fit to an effective
Ivantsov solution by calculating the parabolic tip radius
$\rho_{tip}^p$ from the slope of z versus $x^2$ in the region behind
the tip where the curve becomes straight. The tip radius obtained
this way is $\rho_{tip}^p=3.84$ and the tip Peclet number
\begin{equation}
Pe^{\rm p}=\frac{V_{tip} \rho_{tip}^{\rm p}}{2D} = 0.237
\label{Hyper}
\end{equation}
The relative difference between the $Pe^{\rm p}$ and $Pe^{\rm IVAN}$
(0.257) \cite{Beckermann} obtained from the Ivantsov relation is less
than 8$\%$ percent. Thus, using $\rho_{tip}^{\rm p}$ does indeed remove
uncertainties from using the value at just one point.

\ifXXX
\begin{figure}[htb]
 \centering
 \includegraphics[width=\figwidth]{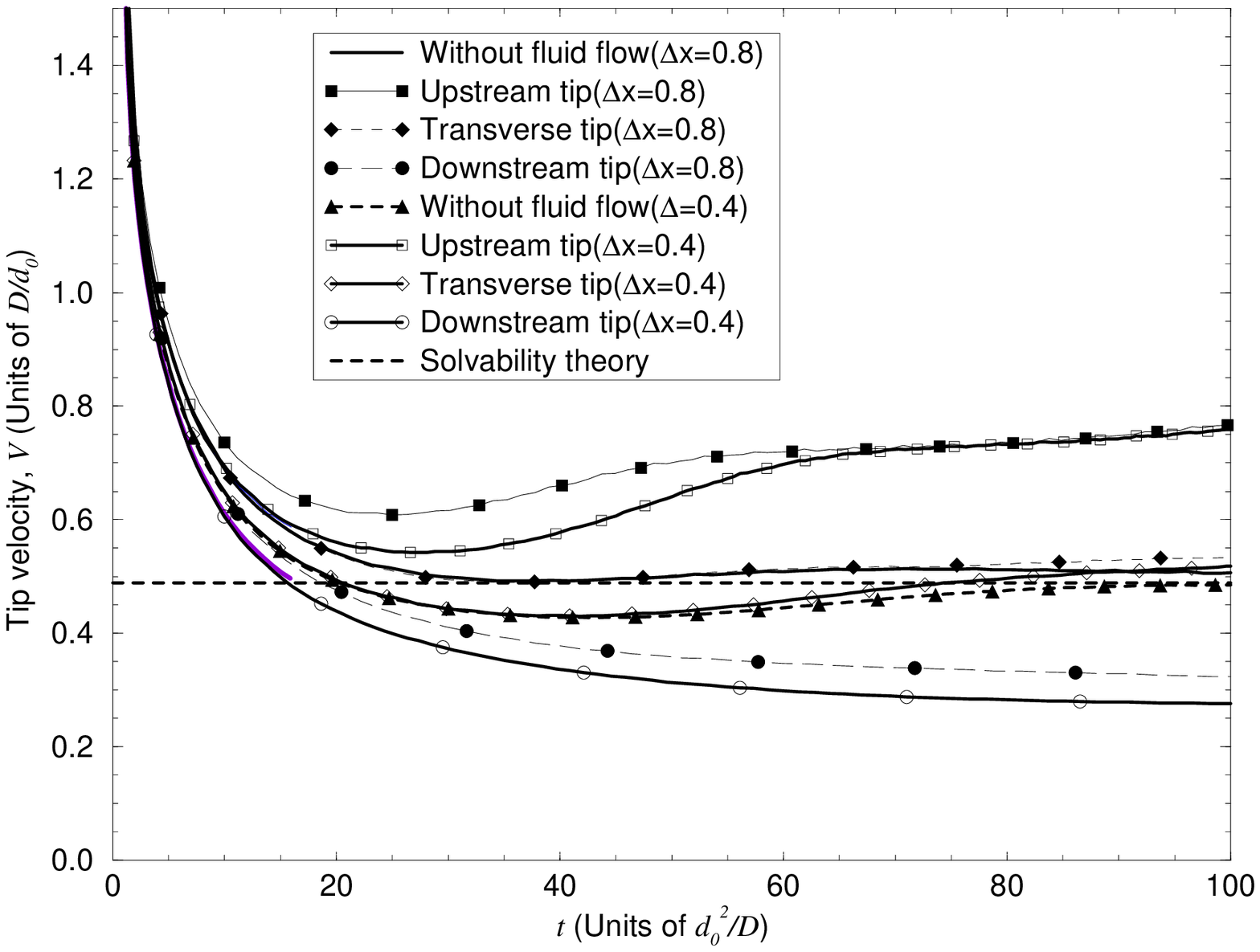}
 \includegraphics[width=\figwidth]{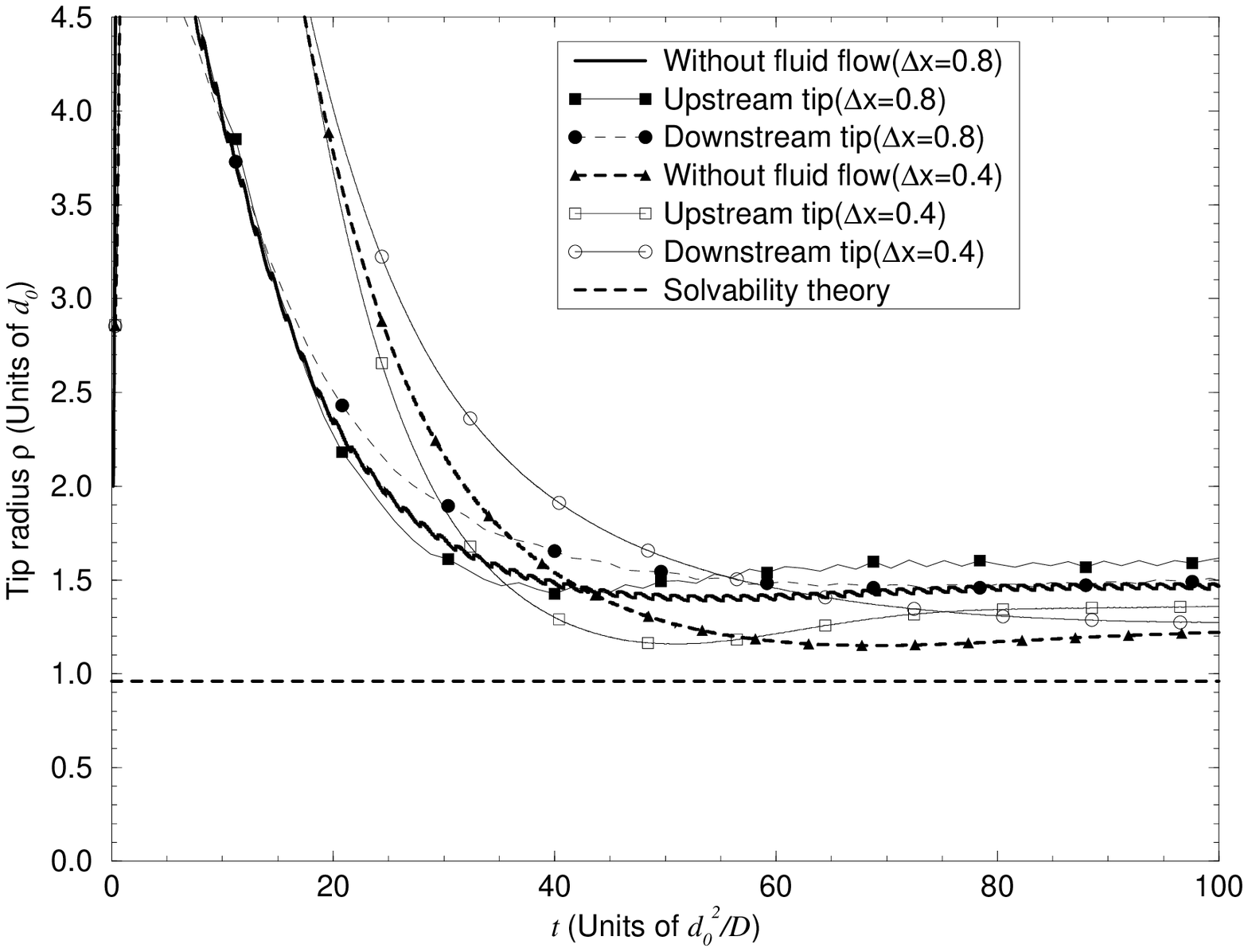}
\caption{Tip velocities and tip radii in 2D}
\label{tipvel2D}
\end{figure}
\fi

\ifXXX
\begin{figure}[htb]
 \centering
 \includegraphics[width=\figwidth]{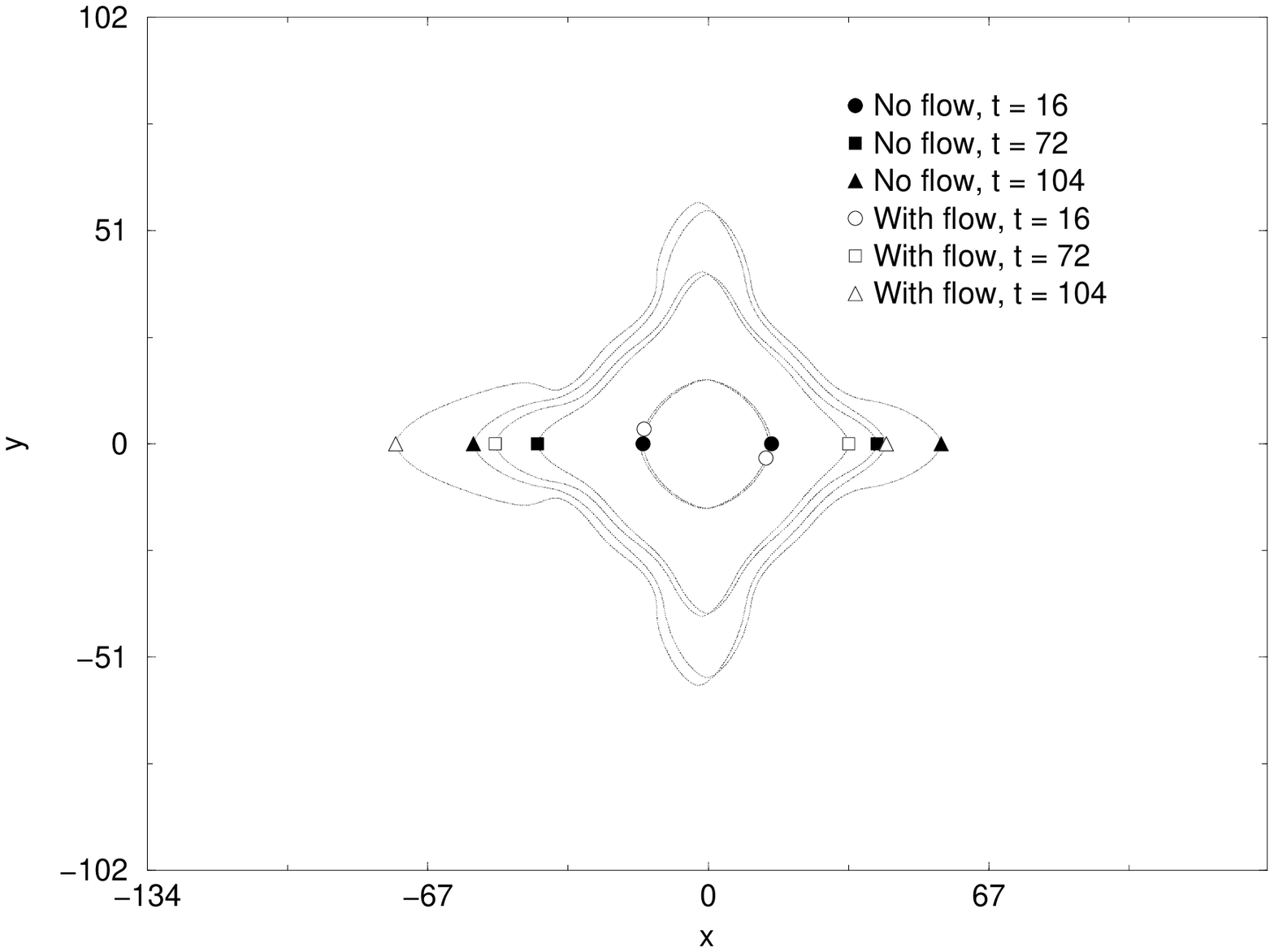}
\caption{Computed interfaces in 2D without and with fluid flow at
t=16, 72, and 104}
\label{2Dinter}
\end{figure}
\fi

Figures \ref{tipvel2D} and \ref{2Dinter} and Table \ref{total}
summarize the results for 2D simulations. For the upstream tips, the
steady tip velocity for $\Delta x$=0.8 and $\Delta x$=0.4 agree with
the results of Beckermann et al.\cite{Beckermann} within 9$\%$
percent.  For $\Delta x$=0.4, the upstream tip grows approximately
55$\%$ faster than in the growth without flow at t=100, while the
downstream tip grows approximately 45$\%$ slower. Beckermann et al.
reported that the upstream tip grows 40$\%$ faster and the downstream
tip grows more than 30$\%$ slower.  The trend for changing the tip
velocities according to the growing directions agrees well with the
results by Beckermann et al. \cite{Beckermann} For the upstream tips,
the ratios $\rho_{tip} /\rho_{tip}^0$ with and without fluid flow are
1.09 and 1.12 for $\Delta x$=0.8 and 0.4, respectively. Those values
also agree well with the result (1.11) obtained by Beckermann et al.
\cite{Beckermann} The ratio $\sigma^*_0 /\sigma^*$, where the
subscript $0$ refers to the the solution in the absence of fluid flow and
$\sigma^*=2d_0D/{\rho_{tip}}^2V_{tip}$, are 1.95 and 0.63 for the
upstream tip and downstream tip, respectively.  These results agree
reasonably well with those of Beckermann et al.

\ifXXX
\begin{table}[htb]
\caption{Results for the simulations of dendrite growth in 2D and 3D}
\label{total}
\begin{center}
\begin{tabular}{|l||c|c|c||c|}
 \hline
		   & Solvability solution(2D) & 2D($\Delta$x=0.8) & 2D($\Delta$x=0.4) & 3D($\Delta$x=0.8)\\
\hline
\hline
$V_{tip}$(no flow)            & 0.489 & 0.502 & 0.486 & 0.915\\
\hline
$V_{tip}^{upstream}  $(flow)&       & 0.766 & 0.761 & 1.038\\
\hline
$V_{tip}^{downstream}$(flow)&       & 0.324 & 0.276 & 0.800\\
\hline
$V_{tip}^{transverse}$(flow)&       & 0.533 & 0.516 & 0.915\\
\hline
$\rho_{tip}$(no flow)          & 0.959 & 1.47  & 1.22  & 2.90 \\
\hline
$\rho_{tip}^{upstream}$(flow)&       & 1.60  & 1.36  & 3.50 \\
\hline
$\rho_{tip}^{downstream}$(flow)&     & 1.50  & 1.28  & 2.41 \\
 \hline
\end{tabular}
\end{center}
\end{table}
\fi

A detailed comparison is not meaningful because
our simulations have not reached steady state, and in some cases are not 
fully grid converged, as shown in Table \ref{total}.
Figure \ref{2Dinter} shows interfaces of the dendrites with and 
without fluid flow in 2D. The interfaces are in good agreement with 
previous results. \cite{Beckermann}

\subsection{3D computations}

\ifXXX
\begin{figure}[htb]
\centering
 \includegraphics[width=\figwidth]{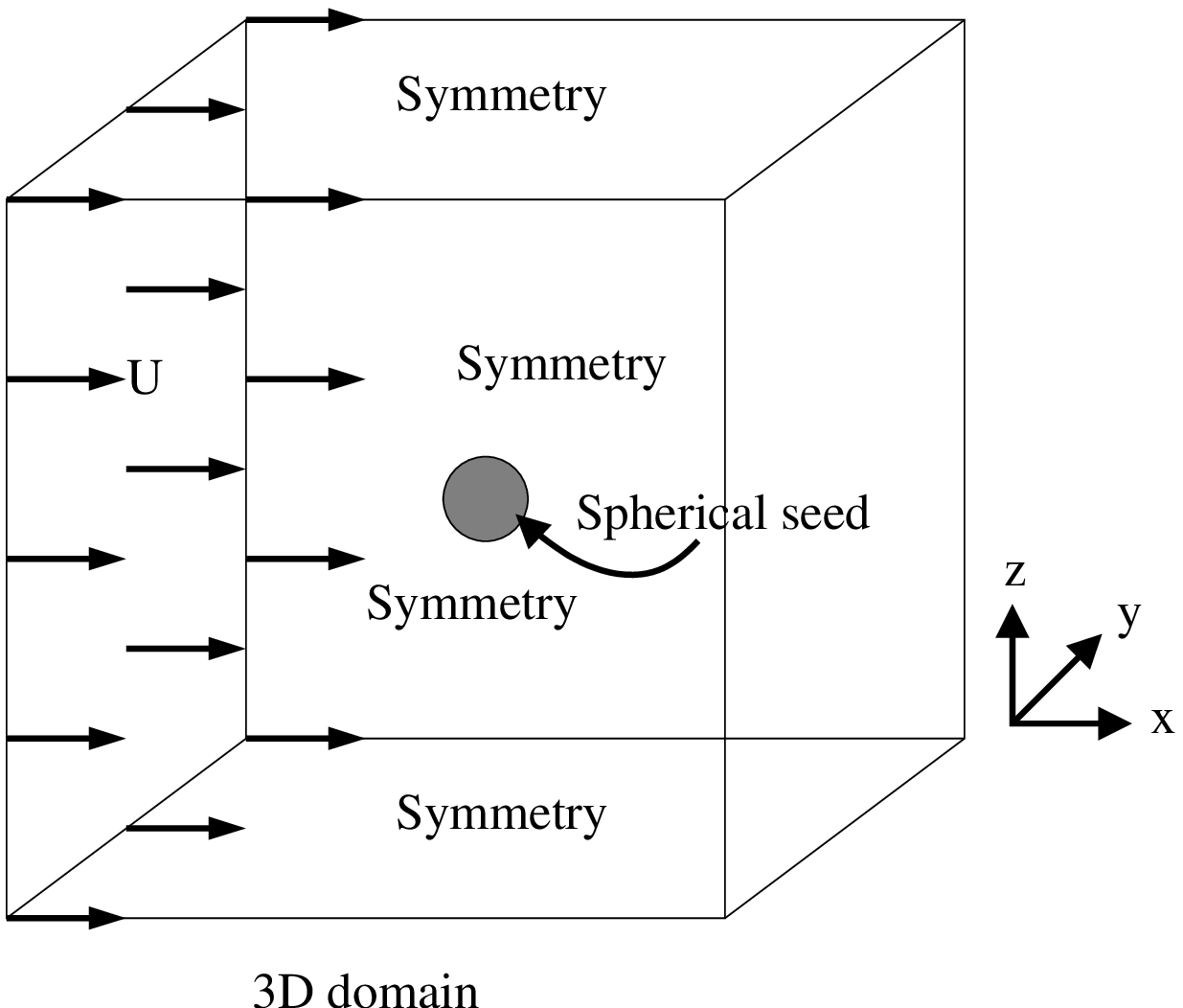}
\caption{Schematic diagram of 3D computation domain and boundary
conditions}
\label{3Ddomain}
\end{figure}
\fi

We simulated 3D dendrite growth with fluid flow using a cube
computation domain with a spherical seed, as shown in Figure
\ref{3Ddomain}. The inlet velocity boundary conditions are imposed on
the left side boundary and their values are $u_x$=U, $u_y$=0, and
$u_z$=0. For this particular problem, we again have $U=1$. The
largest $\Delta x$ is 6.4, while the smallest $\Delta x$ is 0.8.
Time increments of $\Delta t$=0.016 and $\Delta t$=0.008 are used to
model growth without and with fluid flow, respectively.  The other
parameters are identical to the parameters for the 2D analysis in
previous section. Figure \ref{3Dgrid} shows several of the grid
configurations in the analysis. The scale factor $\gamma$ used in the
adaptive grid procedure was 10 for this simulation.

\ifXXX
\begin{figure}[htbp]
 \centering
 \includegraphics[height=4in,angle=270]{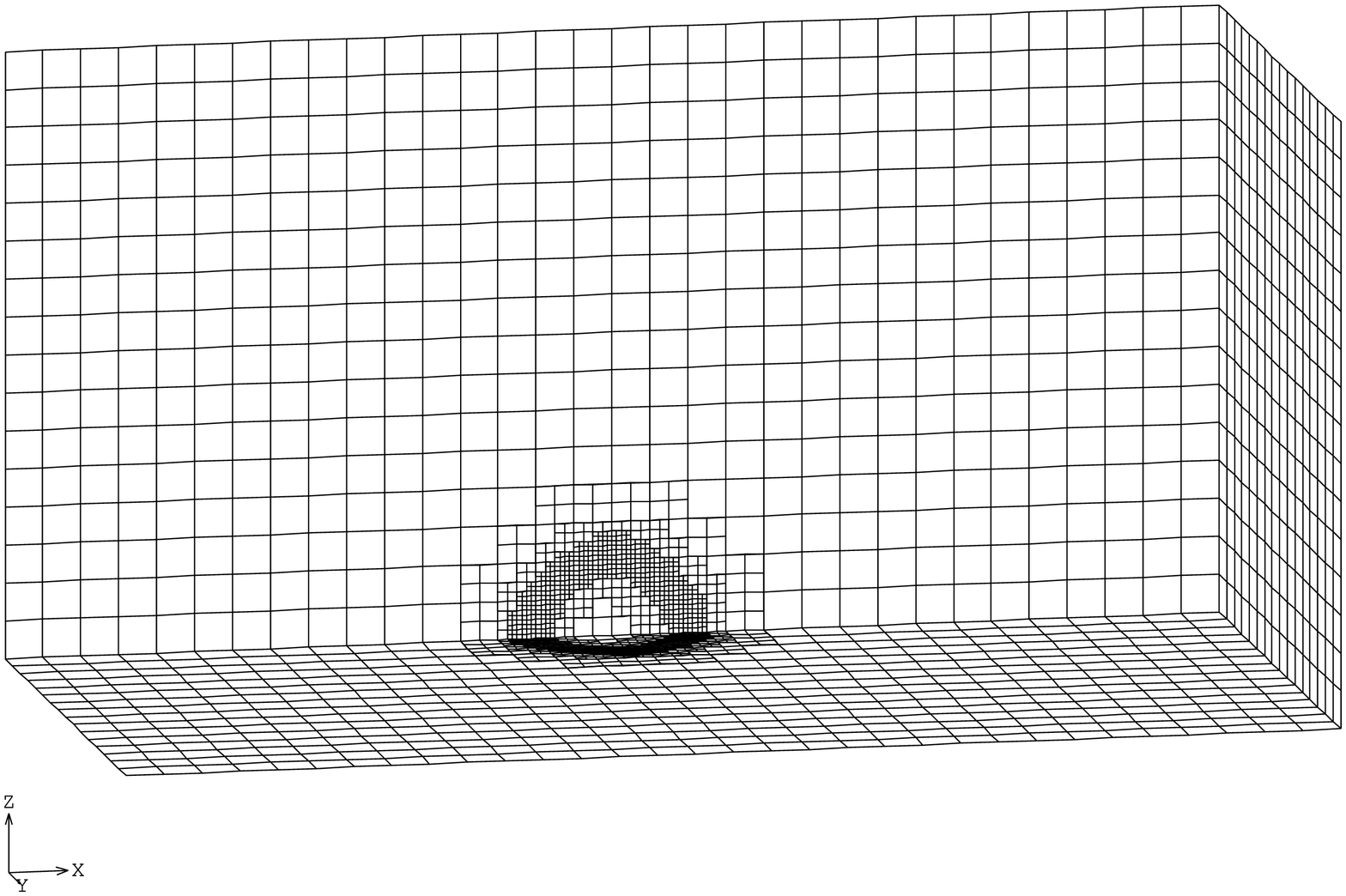}
 \includegraphics[height=4in,angle=270]{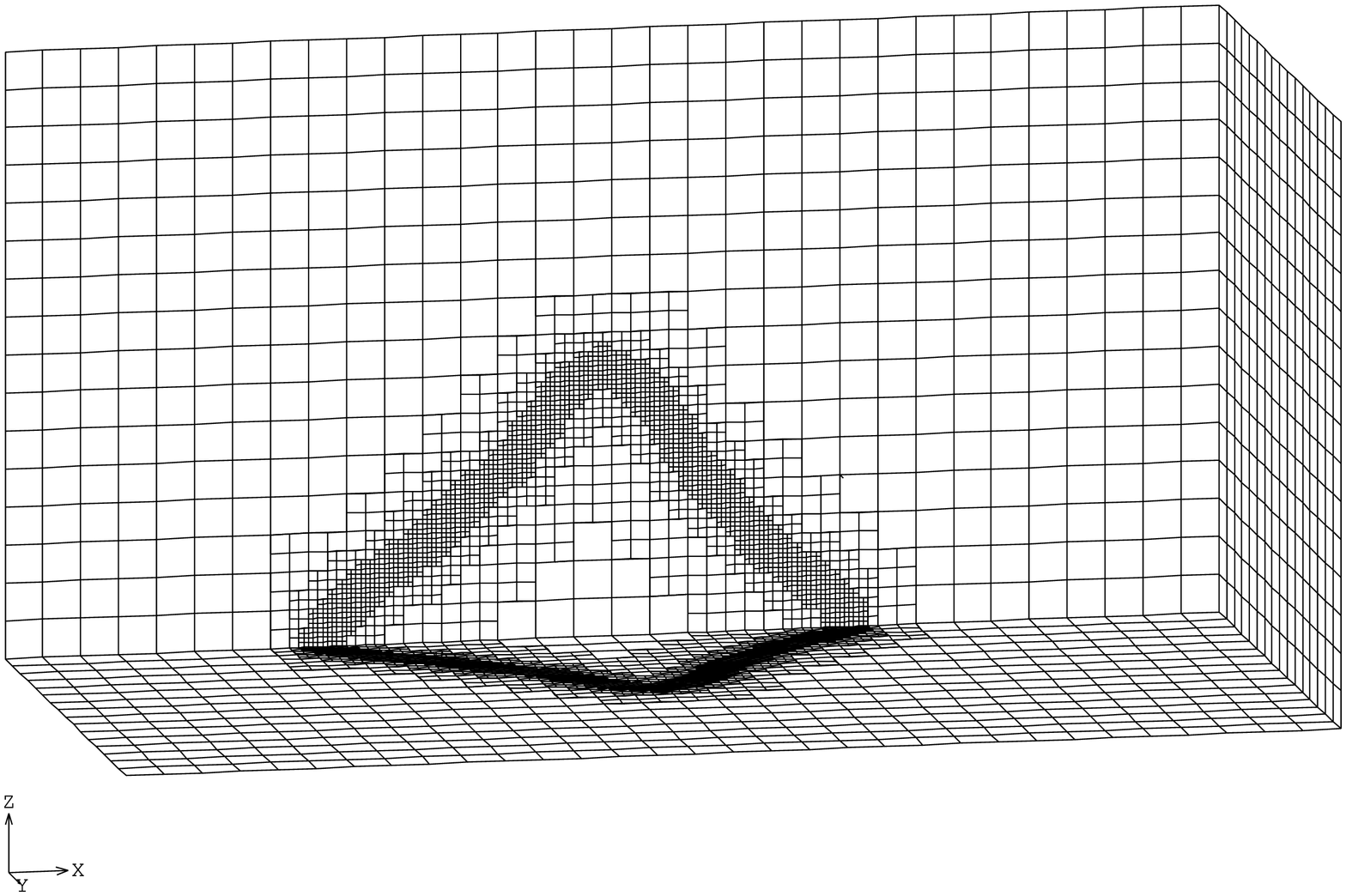}
 \includegraphics[height=4in,angle=270]{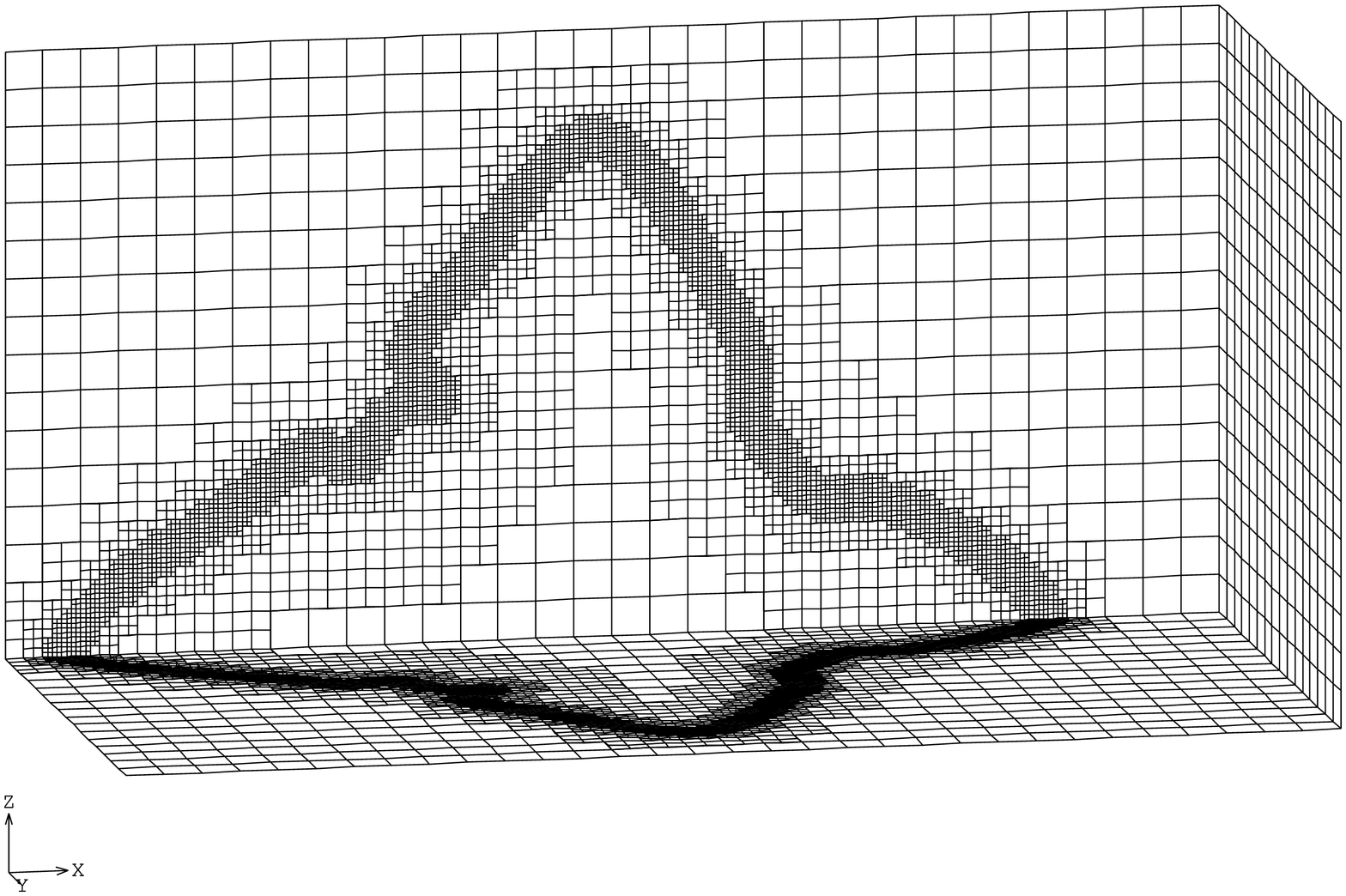}
\caption{Grid configurations of adapted grids t=9.6, 43.2, and 86.4}
\label{3Dgrid}
\end{figure}
\fi

%

The simulations using $\Delta$x=0.8 and $\Delta$x=0.4 consumed about
40 hours and 190 hours, respectively, on a single processor of the
IBM RS/6000 machine with clock rate of 200MHz. In the simulation
where $\Delta x = 0.8$, the initial mesh consisted of 10,607
elements, and the final mesh had 208,357 elements.  These numbers
should be compared with a fully dense mesh, which would have had over
$4\times 10^6$ elements. With fluid flow, computing the velocity and
pressure fields consumes approximately 80-90$\%$ of the total CPU
time.  We performed some the simulations using a different time
increment for velocity field $\Delta t_V$, larger than the time
increment for temperature and $\phi$ fields $\Delta t_{ \theta
\phi}$.  For $\Delta x$=0.8 and $\Delta t_V$=5$\Delta t_{\theta
\phi}$, the CPU time decreased by 76$\%$, and the results for
$\rho_{tip}$ and $V_{tip}$ agree within 5$\%$ with the results
obtained by using $\Delta t_V$=$\Delta t_{\theta \phi}$.  The
detailed results presented next were computed using $\Delta
t_{V}$=$\Delta t_{\theta \phi}$=0.008 in order to compare our results
with the previous 2D results. This run took approximately 250 hours
on 16 processors of the ORIGIN2000 at the National Center for
Supercomputing Applications (NCSA).  The same run using $\Delta
t_{\theta \phi}$=0.016 and $\Delta t_V$=5$\Delta t_{\theta \phi}$
consumed approximately one eighth of the run time for $\Delta
t_V$=$\Delta t_{\theta \phi}$ =0.008, and the deviation in the
results was less than 3$\%$. Thus, a run time of about 32 hours on 16
processors yielded essentially the same result.

Without fluid flow, the steady scaled tip velocity $V_{tip} d_0/D$ is
0.0318 and the ratio $V_{tip}^{2D} / V_{tip}^{3D}$ for $\Delta$x=0.8
is 0.55. Karma and Rappel\cite{Karma+Rappel} reported that this ratio
is 0.39 for $\Delta$=0.65 and effective anisotropy
$\epsilon_e$=0.0269.  The $\rho_{tip}/ d_0$ is 20.86 and the ratio
$\rho_{tip}^{2D} / \rho_{tip}^{3D}$ for $\Delta$x=0.8 is 0.51. The
ratio $\sigma^*_{2D} /\sigma^*_{3D}$ is 7.09, where $\sigma^*$ is the
selection constant $\sigma^*=2Dd_0/\rho_{tip}^2V_{tip}$.

\ifXXX
\begin{figure}[ht]
 \centering
 \includegraphics[width=\figwidth]{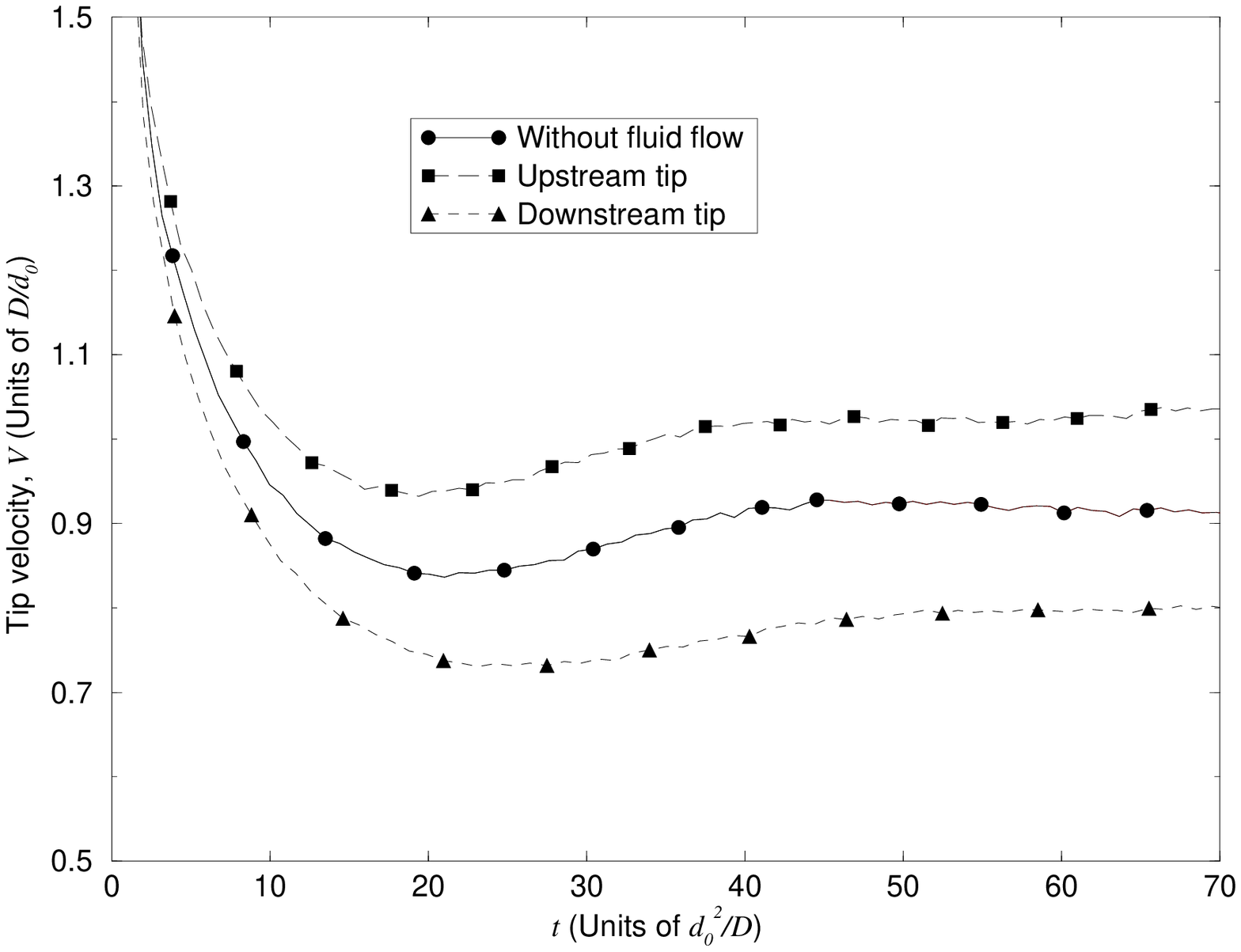}
 \includegraphics[width=\figwidth]{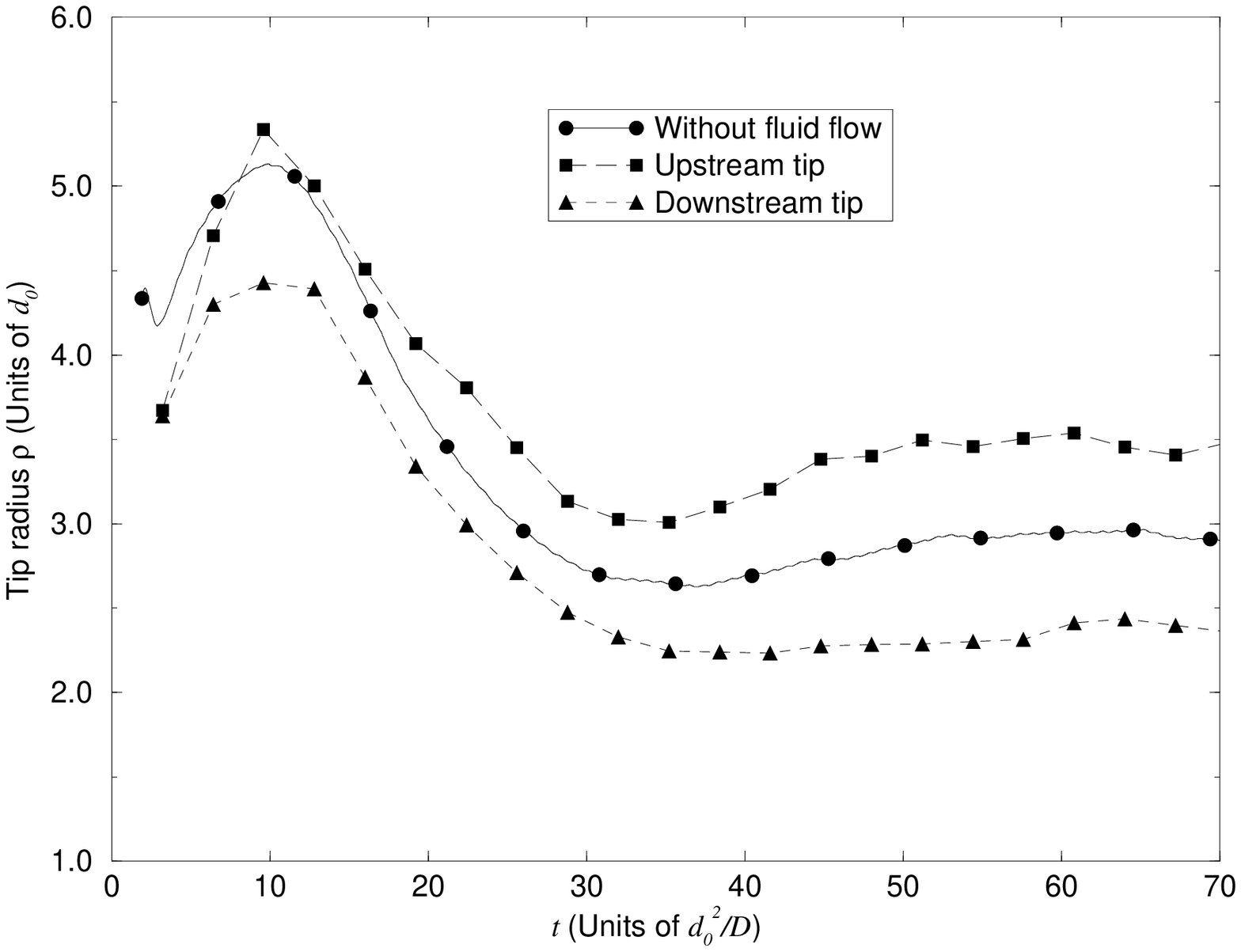}
\caption{Tip velocities and tip radii in 3D}
\label{tipvel3D}
\end{figure}
\fi

The time evolution of the dendrite tip velocities and tip radii are
shown in Figure \ref{tipvel3D}.  With fluid flow, the upstream tip
grows 13$\%$ faster than it does without flow, while the downstream
tip grows 13$\%$ slower.  We stopped the calculations at t=86.5, before
the diffusion field encounters the end of the computational domain.
These results seem to show that the effect of 3D flow on the upstream
dendrite tip velocity is much weaker than the effect of 2D flow.
However, the reason for the trend is that the effect of forced flow
on tip velocity is highly dependent on the ratio $U/V_{tip}$ where U
is the inlet velocity. The ratio $U/V_{tip}^{3D}$ is 45$\%$ smaller
than the ratio $U/V_{tip}^{2D}$, and this accounts for the reduced
effect.  The ratio $\rho_{tip}/\rho_{tip}^0$ is 1.21 for 3D for the
upstream tip, which is slightly larger than the ratio
$\rho_{tip}/\rho_{tip}^0$=1.09 for 2D.  The tip radius for the
downstream in 3D decreases by 17$\%$ due to the flow effect, while
the tip radius in 2D increases by 2$\%$.  Thus, the effect of the
forced flow on the downstream tip is much stronger in 3D than in 2D.
The cause of the trend is that in 2D the fluid flows vertically over
the dendrite, while in 3D the fluid flows both vertically and
laterally over the dendrite.

Figure \ref{3Dinter} shows several interfaces in the $x$-$z$ plane. The
tilt angle of the transverse arms into the flow in 3D is $2.5^\circ$,
while the tilt angle in 2D is $2.3^\circ$.  From Figure \ref{3Dinter}
we can see that sidebranches begin to appear on the upstream side of
the transverse tip more quickly than on the other sides.

\ifXXX
\begin{figure}[htb]
 \centering
 \includegraphics[width=\figwidth,angle=0]{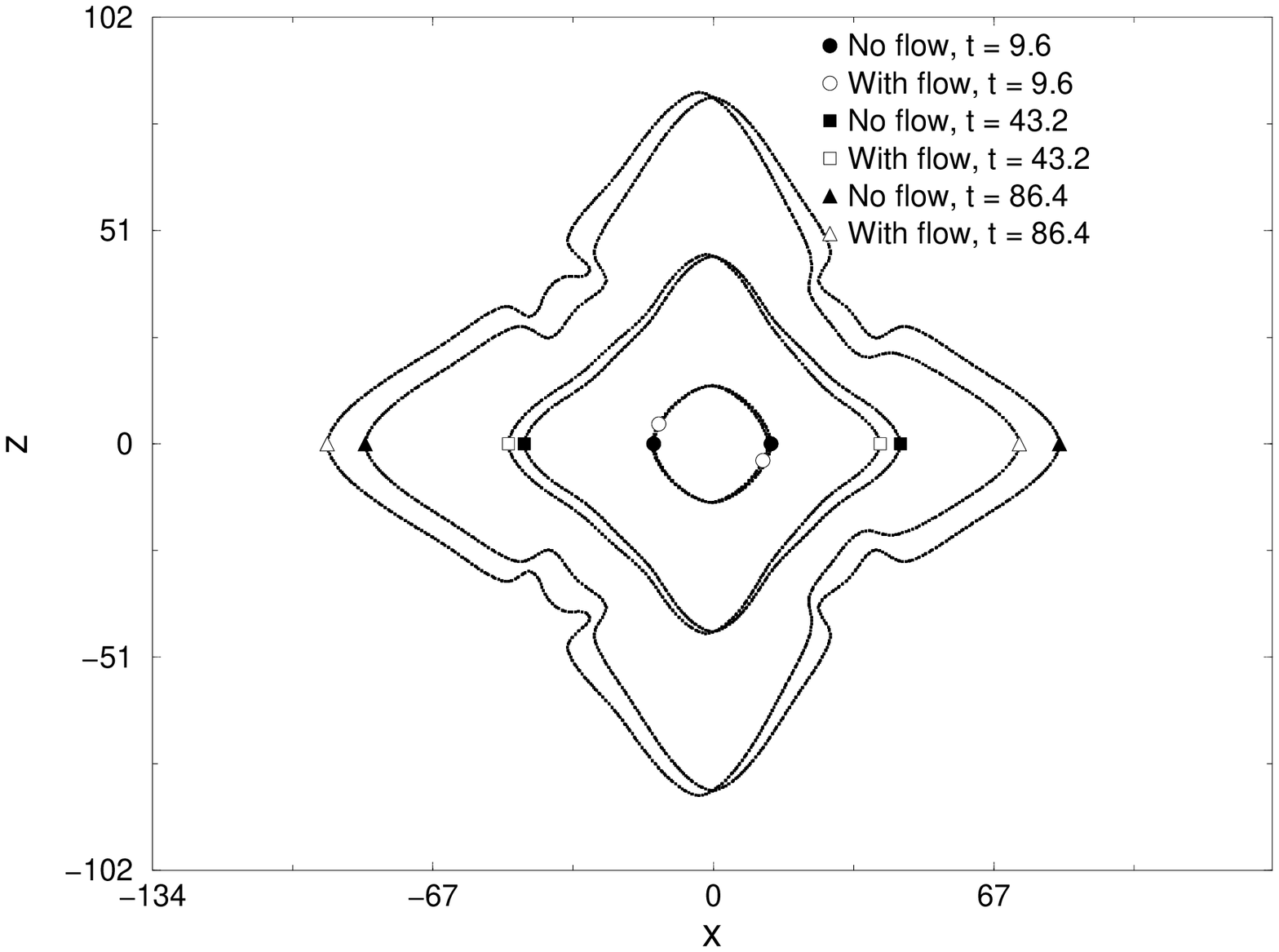}
\caption{Computed interfaces in 3D without and with fluid flow at
t=9.6, 43.2, and 86.4}
\label{3Dinter}
\end{figure}
\fi

Bouissou et al.\cite{bouissou} showed that when surface tension
effects are neglected, the Peclet number $Pe_{V}$ computed using $V_{tip}$
is related to the Peclet number computed using the inlet
velocity $Pe_{U}$ for a given undercooling.  In order to assess the
effect of the forced flow on the tip Peclet number, We computed the
the difference between the tip Peclet number with fluid flow
$Pe_{V}^{f}=V^f\rho^f/2D$ and the tip Peclet number without fluid
flow $Pe_{V}^{0}=V^0\rho^0/2D$ , divided by the Peclet number related
to the inlet velocity $Pe_U^{f}=U\rho^f/2D$
\begin{equation} \Delta Pe=\frac{Pe_{V}^{f}-Pe_{V}^{0}}{Pe_U^{f}}
\label{deltaPe}
\end{equation}
We found that for the upstream tip in 2D, $Pe_{U}^{2D}$ is 0.305, and
it is similar to $Pe_{U}^{3D}$(=0.275), even though the effect of the
forced flow on the tip velocities and tip radii in 2D and 3D is much
different.

We found that the ratio $\sigma^*_0 /\sigma^*$ for 3D is 1.675 for
upstream tip and 0.603 for the downstream tip.  With flow present, it
takes somewhat longer for $\sigma^*$ to settle onto a steady value,
as seen in Figure \ref{sigma3D}.  The trend in $\sigma^*$ for the
upstream tip with fluid flow agrees with experiments by Bouissou et
al.\cite{bouissou} on alloys of PVA and ethanol, while the trend is
opposite to the experiments by Lee et al.\cite{lee} on SCN forced
past a needle.  Further work is needed to examine this phenomenon.

\ifXXX
\begin{figure}[htb]
 \centering
 \includegraphics[width=\figwidth]{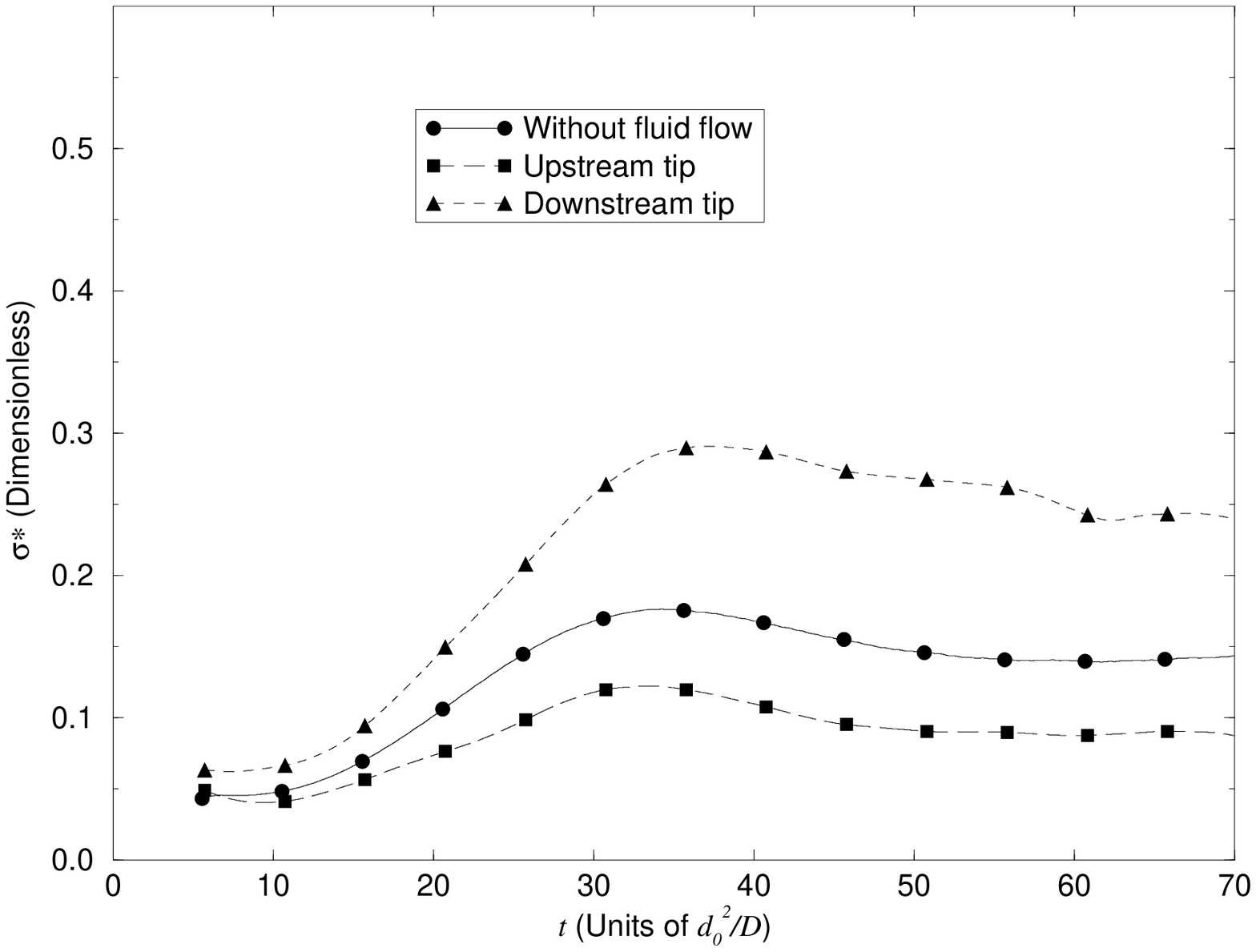}
\caption{$\sigma^*$s in 3D}
\label{sigma3D}
\end{figure}
\fi

\section{Conclusions}

In this paper, we presented an efficient algorithm using 3D adaptive
grid refinement in order to study the effect of fluid flow on 3D
dendrite growth using a phase-field model. We simulated 3D dendrite
growth with fluid flow using adapted grids with at most 300,000
elements, while over 4,000,000 elements would be required for an
uniform grid.  We also present a parallel implementation by using
Charm++ FEM framework.  A speed-up factor SP=28.8 for 32 processors
is typical for the larger meshes.

We also introduce a new scheme using a hyperbolic tangent function to
interpolate tip position.  We found that the scheme, which requires
just two points for interpolation, has a smaller oscillation
amplitude than third-order polynomial interpolation using 4 points.

Test cases showed that the dendrite tip velocities and radii for 2D
dendrite growth were in satisfactory agreement with solvability
theory and previous computational results\cite{Beckermann}.  We found
that the effect of fluid flow on dendrite tip velocity in 2D is much
larger than in 3D, because the ratio $U/V_{tip}^{2D}$ is
approximately two times larger than the ratio $U/V_{tip}^{3D}$.  For
the downstream tip, the tip radius in 3D decreases by 17$\%$, and the
tip radius in 2D increases by 2$\%$.  The cause of the trend is that
in 3D the fluid flows both vertically and laterally over the dendrite
and the effect of the forced flow in 3D is much stronger than in 2D
where the fluid flows vertically over the dendrite.

We examined the effect of the forced flow on the tip Peclet number by
computing $\Delta Pe$: the difference between the tip Peclet number
in the growth with fluid flow and the tip Peclet number in the growth
without the fluid flow, divided by the Peclet number related to the
forced flow.  We found that $\Delta Pe$ in 2D and in 3D are within
10$\%$, even though the tip radii and velocities vary by much more
than that.  In 3D growth with fluid flow, ${\sigma^*}$ for the
upstream tip decreases by 39$\%$.  The trend for ${\sigma^*}$ for the
upstream tip agrees with experiments (reduction of 37$\%$) by
Bouissou et al.\cite{bouissou} on alloys of PVA and ethanol.  The
ratio ${\sigma^*_0 /\sigma^*}^{3D}$ for the upstream tip is 14$\%$
smaller than the ratio ${\sigma^*_0 /\sigma^*}^{2D}$.

\section*{Acknowledgments}

This work has been supported by the NASA Microgravity Research
Program, under Grant NAG8-1249, and the National Science Foundation,
under grant NSF DMS 98-73945. The authors wish to thank S.~Kale,
M.~Bhandarkar, O.~Lawlor and T.~Hinrichs for their help with the
parallelization of the code.  We also thank C.~Beckermann and
A.~Karma for helpful discussions and sharing of their results.

\ifDRAFT
 \bibliographystyle{revtex}
 \bibliography{jeong}

\begin{thebibliography}{10}

\bibitem{mullins}
W.~W. Mullins and R.~F. Sekerka.
\newblock J. Appl. Physics {\bf 35}, 444 (1964).

\bibitem{LanII80}
J.~Langer.
\newblock ``Lectures in the Theory of Pattern Formation", in {\em Chance and
  Matter}, Les Houches Session XLVI, edited by J. Souletie, J. Vannenimus and
  R. Stora, North Holland, Amsterdam  629--711 (1987).

\bibitem{Kes88}
D.~A. Kessler, J.~Koplik and H.~Levine.
\newblock Adv. Phys. {\bf 37}, 255 (1988).

\bibitem{Ben84}
E.~Ben-Jacob, N.~Goldenfeld, B.~Kotliar and J.~Langer.
\newblock Phys. Rev. Lett. {\bf 53}, 2110 (1984).

\bibitem{KessI84}
D.~Kessler, J.~Koplik and H.~Levine.
\newblock Phys. Rev. A {\bf 30}, 3161 (1984).

\bibitem{Karma+Rappel}
A.~Karma and W.-J. Rappel.
\newblock Phys. Rev. E. {\bf 53}, 3017 (1995).

\bibitem{Provatas1998}
N.~Provatas, N.~Goldenfeld and J.~Dantzig.
\newblock Phys. Rev. Lett. {\bf 80}, 3308 (1998).

\bibitem{Tae}
Y.-T. Kim, N.~Goldenfeld and J.~Dantzig.
\newblock Phys. Rev. B {\bf 62}, 2471 (2000).

\bibitem{osher}
S.~Chen, B.~Merriman, S.~Osher and P.~Smereka.
\newblock J. Comp. Phys. {\bf 135}, 8 (1997).

\bibitem{davis}
S.~H. Davis.
\newblock {\em Theory of Solidification\/} (Cambridge University Press, 2001).

\bibitem{Dantzig}
J.~A. Dantzig and L.~S. Chao.
\newblock In {\em Proc. 10th U.S. Nat. Cong. of Appl. Mech.\/}, edited by
  J.~Lamb,  249 (1986).

\bibitem{Provatas1999}
N.~Provatas, N.~Goldenfeld and J.~Dantzig.
\newblock J. Comp. Phys {\bf 148}, 265 (1999).

\bibitem{Beckermann}
C.~Beckermann, H.-J.Diepers, I.~Steinbach, A.~Karma and X.~Tong.
\newblock J. Comp. Phys. {\bf 154}, 468 (1999).

\bibitem{Gresho1995}
P.~M. Gresho, S.~T. Chan, M.~A. Christon and A.~C. Hindmarsh.
\newblock Int. J. Numer. methods fluids {\bf 21}, 837 (1995).

\bibitem{Mikic}
X.~Mikic and E.~C. Morse.
\newblock J. comp. Phys. {\bf 61}, 154 (1985).

\bibitem{Brooks}
A.~N. Brooks and T.~J.~R. Hughes.
\newblock Comp. Methods Appl. Mech. Eng. {\bf 32}, 199 (1982).

\bibitem{charm}
L.~V. Kale and S.~Krishnan.
\newblock In {\em Parallel Programming using C++\/}, edited by G.~V. Wilson and
  P.~Lu,  175 (MIT Press, 1996).

\bibitem{fem-hipc}
M.~Bhandarkar and L.~V. Kal\'{e}.
\newblock In {\em Proceedings of the International Conference on High
  Performance Computing (HiPC)\/} (December 2000).

\bibitem{metis}
G.~Karypis.
\newblock {\em http://www-users.cs.umn.edu/$\,\tilde{ }\:$karypis/metis/\/}.
\newblock Univ.~of Minnesota (2000).

\bibitem{Wheeler}
A.~A. Wheeler, B.~T. Murray and R.~J. Schaefer.
\newblock Physica D {\bf 66}, 243 (1993).

\bibitem{bouissou}
P.~Bouissou, B.~Perrin and P.~Tabeling.
\newblock Phys. Rev. A {\bf 40}, 509 (1989).

\bibitem{lee}
Y.-W. Lee, R.~Ananth and W.~N. Gill.
\newblock J. Cryst. Growth {\bf 132}, 226 (1993).

\end{thebibliography}
\else

\fi

\ifDRAFT
\input{figures}
\fi

\ifPRE
\def\colwidth{5in}
\onecolumn

%
%
\ifDRAFT
 \clearpage
\fi
\begin{table}[htb]
\caption{Results for the simulations of dendrite growth in 2D and 3D}
\label{total}
\begin{center}
\begin{tabular}{|l||c|c|c||c|}
\ifDRAFT
 \hline
\fi
		   & Solvability solution(2D) & 2D($\Delta$x=0.8) & 2D($\Delta$x=0.4) & 3D($\Delta$x=0.8)\\
\hline
\hline
$V_{tip}$(no flow)            & 0.489 & 0.502 & 0.486 & 0.915\\
\hline
$V_{tip}^{upstream}  $(flow)&       & 0.766 & 0.761 & 1.038\\
\hline
$V_{tip}^{downstream}$(flow)&       & 0.324 & 0.276 & 0.800\\
\hline
$V_{tip}^{transverse}$(flow)&       & 0.533 & 0.516 & 0.915\\
\hline
$\rho_{tip}$(no flow)          & 0.959 & 1.47  & 1.22  & 2.90 \\
\hline
$\rho_{tip}^{upstream}$(flow)&       & 1.60  & 1.36  & 3.50 \\
\hline
$\rho_{tip}^{downstream}$(flow)&     & 1.50  & 1.28  & 2.41 \\
\ifDRAFT
 \hline
\fi
\end{tabular}
\end{center}
\end{table}

\ifDRAFT
 \clearpage
\fi
\begin{figure}[htb]
\ifDRAFT
 \centering
 \includegraphics[width=\colwidth]{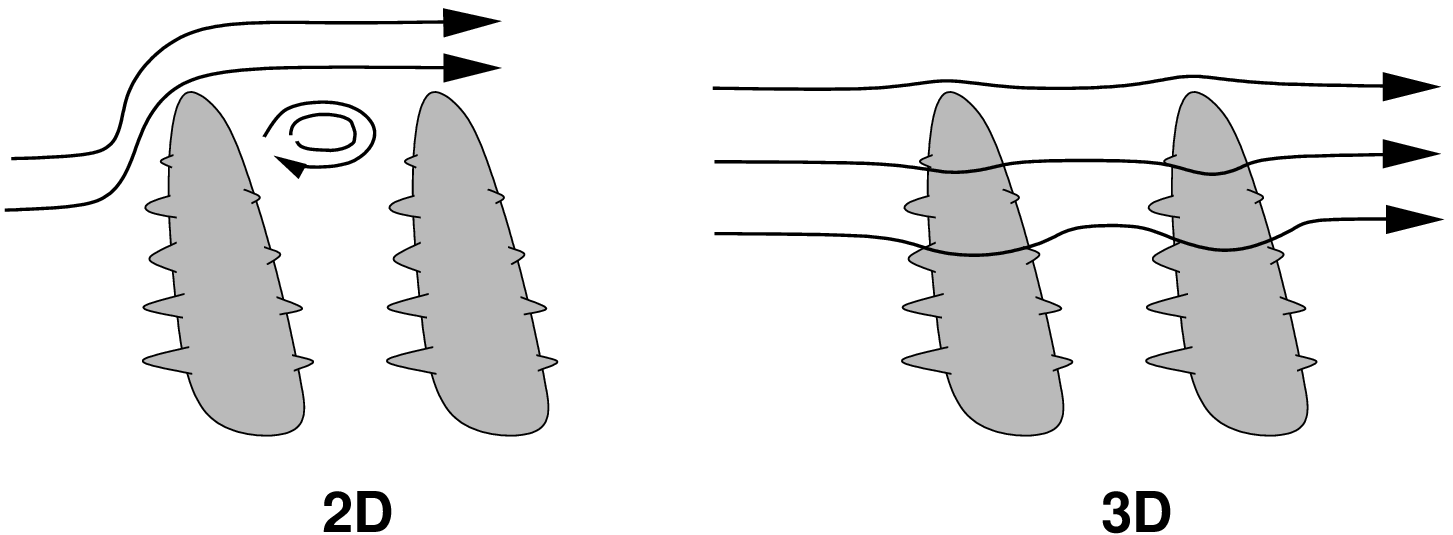}
\fi
\caption{Schematic drawing of the flow over dendrites growing
perpendicular to a superimposed flow, comparing 2D and 3D phenomena.}
\label{2d-3d}
\end{figure}

\ifDRAFT
 \clearpage
\fi
\begin{figure}[htb]
\ifDRAFT
 \centering
 \includegraphics[width=\colwidth]{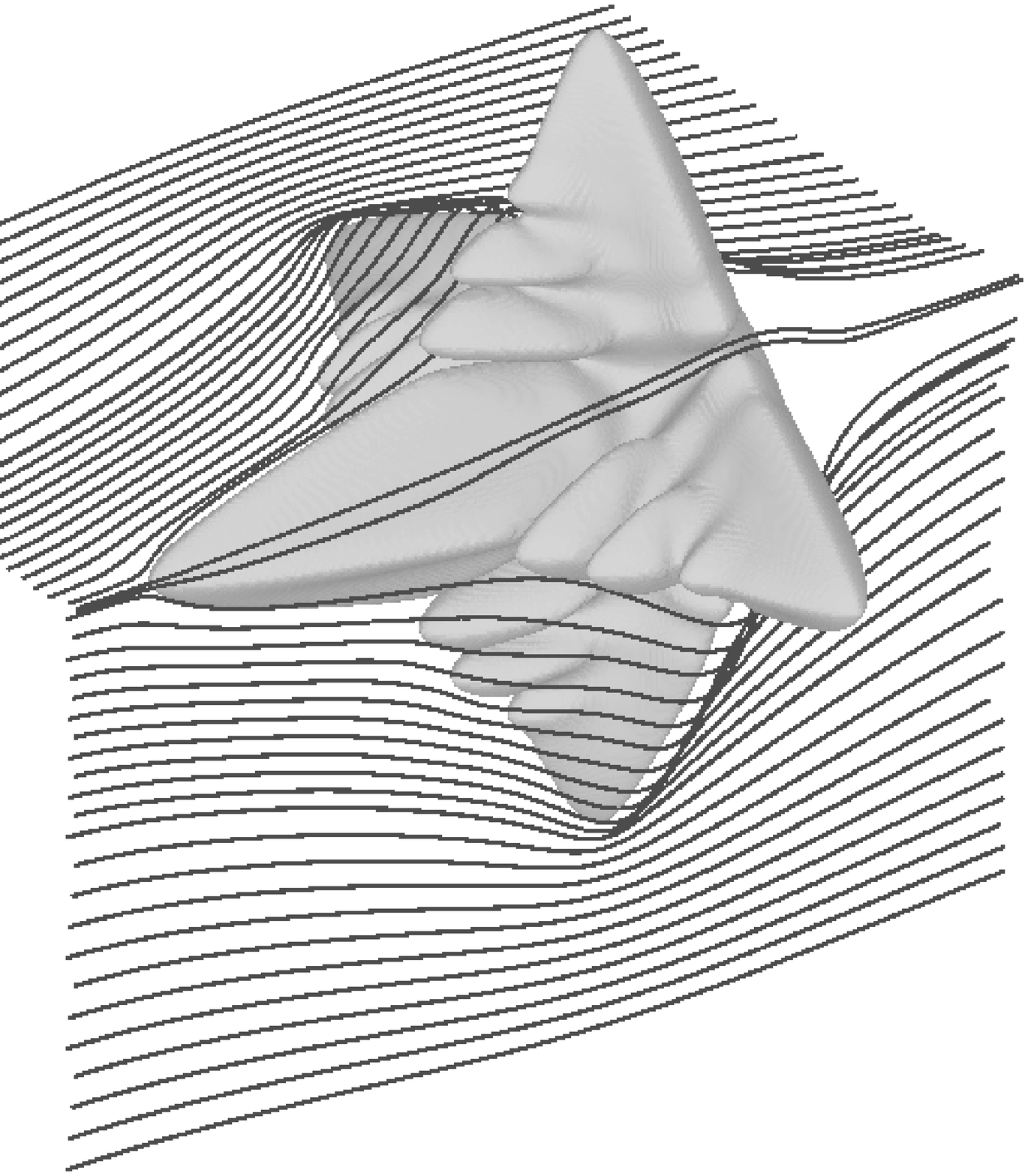}
\fi
\caption{Computed streamtraces for flow over a growing isolated
dendrite.}
\label{dendrite1}
\end{figure}

\ifDRAFT
 \clearpage
\fi
\begin{figure}[htb]
\ifDRAFT
 \centering
\scriptsize{
\begin{verbatim}
type element
  integer                         :: num_element            !! Element number
  integer                         :: level                  !! Refinement level
  integer                         :: lneigh                 !! Number of neighbor elements
  type(connectivity), pointer     :: connect                !! Pointer of connectivity		
  type(connectivity), pointer     :: connect_mid            !! Pointer of connectivity for disconnected nodes
  integer                         :: midindex(6)            !! Index to check for discontinuous nodes
  type(neighbor_elements), pointer:: neighbor               !! Pointer to neighbor elements
  integer                         :: num_parent(LimitLevel) !! Parent element numbers
  integer                         :: num_history(LimitLevel)!! Time step number 
  integer                         :: merge                  !! Index to check if the element should be merged
  real*8                          :: error                  !! Error estimator value at time step n
  integer                         :: nver                   !! Vertex node number for disconnected edge node
  integer                         :: ntype                  !! Element type (liquid/solid/interface)
                                                            !! If phi = -1         , ntype = 1(liquid)
                                                            !! If phi = 1          , ntype = 2(solid)
                                                            !! If -1 < phi < 1     , ntype = 3(interface)
  real*8                          :: pe                     !! Element pressure
  type(element), pointer          :: previous               !! Previous element in linked list
  type(element), pointer          :: next                   !! Next element in linked list
end type element
\end{verbatim}
}
\fi
\caption{Element linked data structure for adaptive grid}
\label{link}
\end{figure}

\ifDRAFT
 \clearpage
\fi
\begin{figure}[htb]
\ifDRAFT
 \centering
 \includegraphics[height=\colwidth,angle=0]{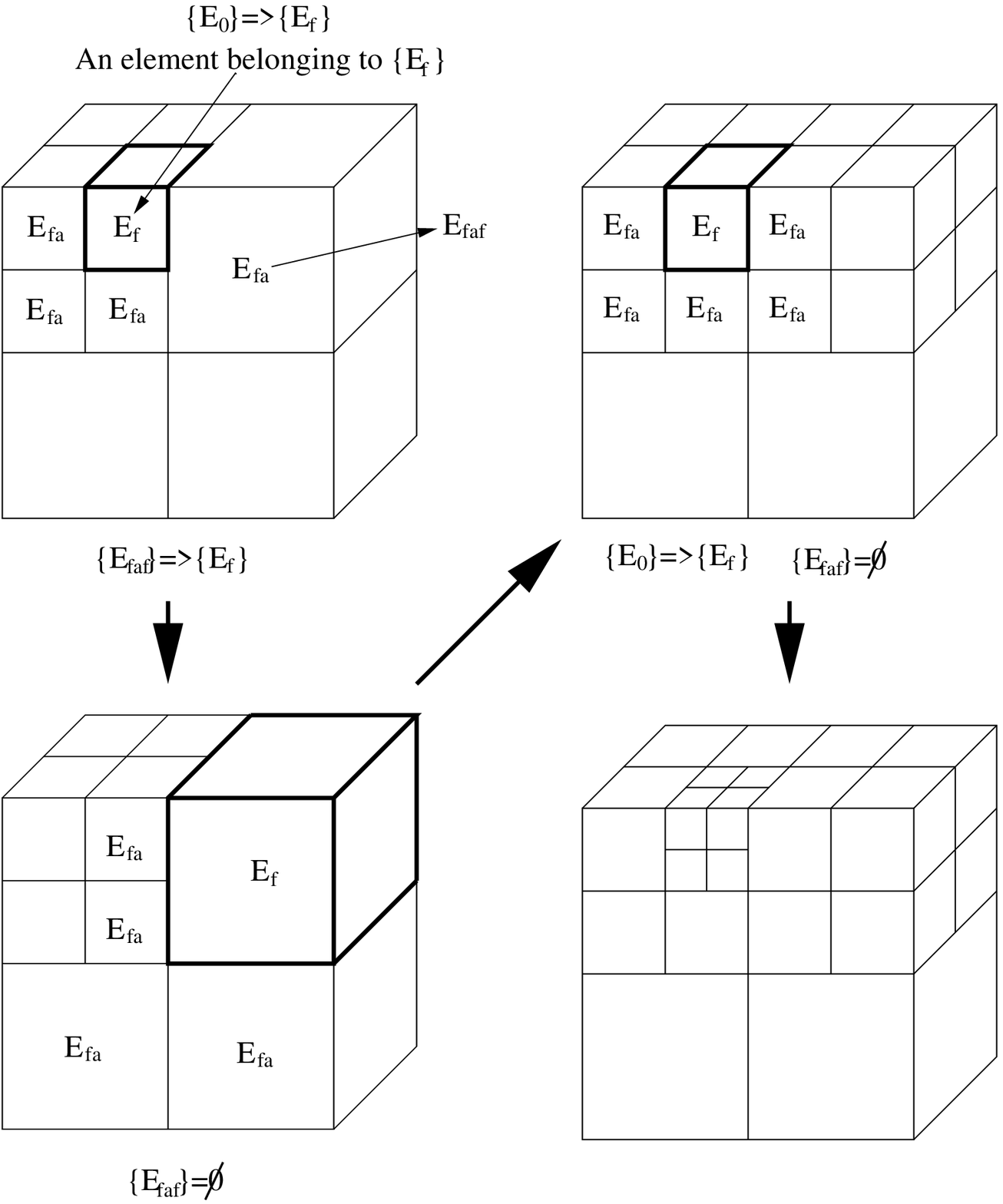}
\fi
\caption{Dividing sequence for refinement. The arrows indicate the
sequence of refinement, as discussed in the text.}
\label{refine}
\end{figure}


\ifDRAFT
 \clearpage
\fi
\begin{figure}[htb]
\ifDRAFT
 \centering
 \includegraphics[width=\colwidth]{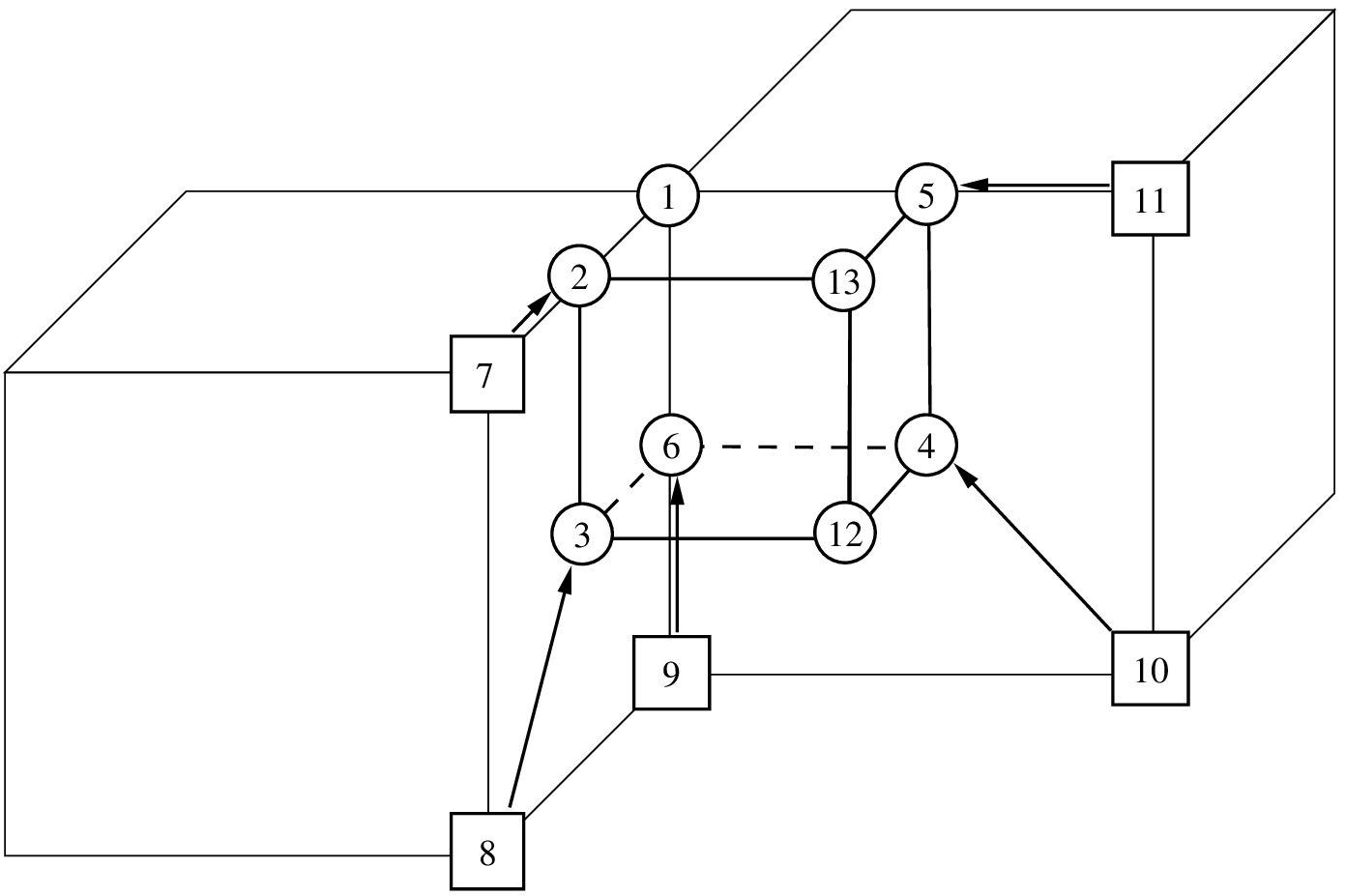}
\fi
\caption{Configurations of disconnected sides and nodes}
\label{discon}
\end{figure}


\ifDRAFT
 \clearpage
\fi
\begin{figure}[htb]
\ifDRAFT
 \centering
 \includegraphics[width=\colwidth]{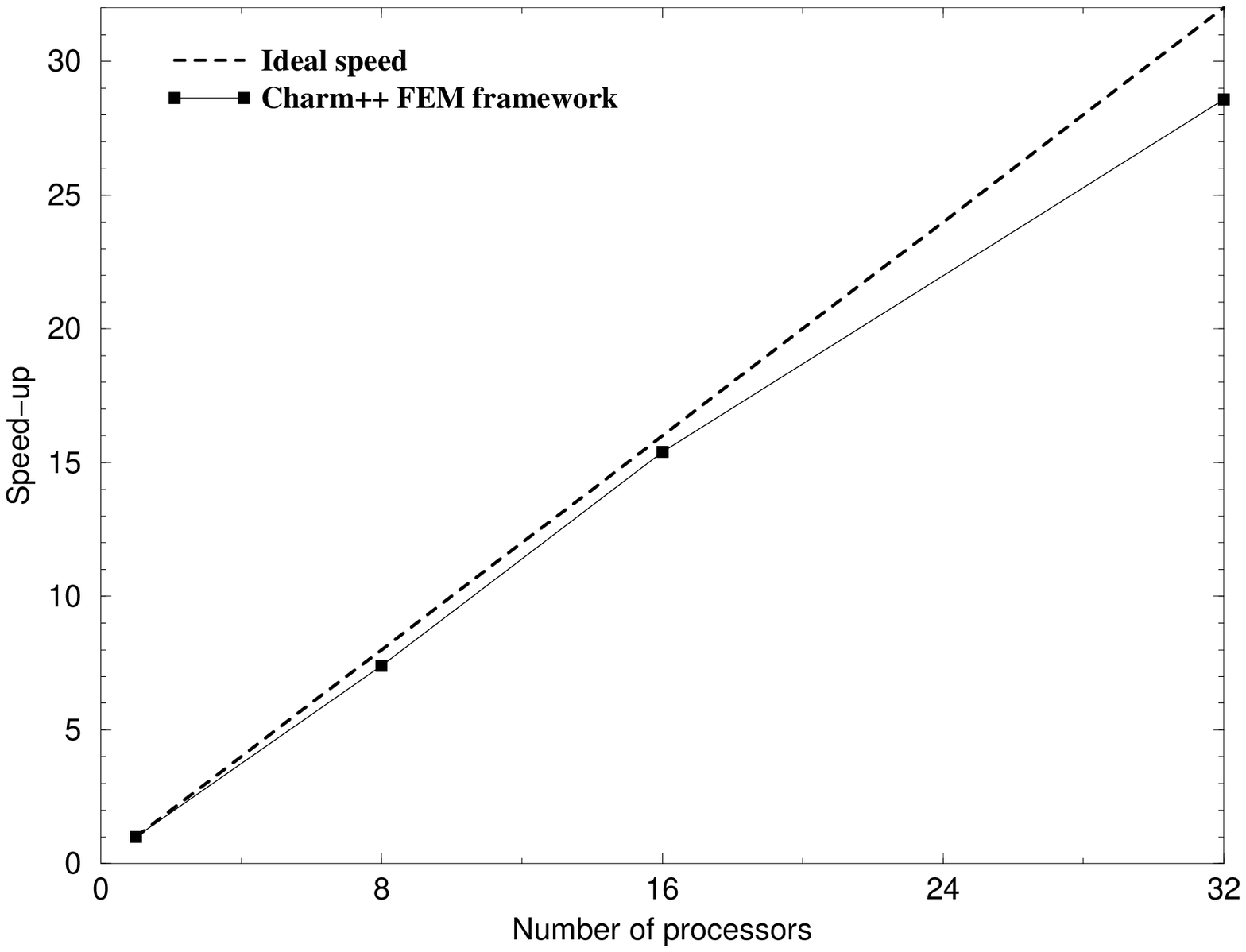}
\fi
\caption{Speed-up for Charm++ FEM framework}
\label{speedup}
\end{figure}

\ifDRAFT
 \clearpage
\fi
\begin{figure}[htb]
\ifDRAFT
 \centering
 \includegraphics[width=\colwidth]{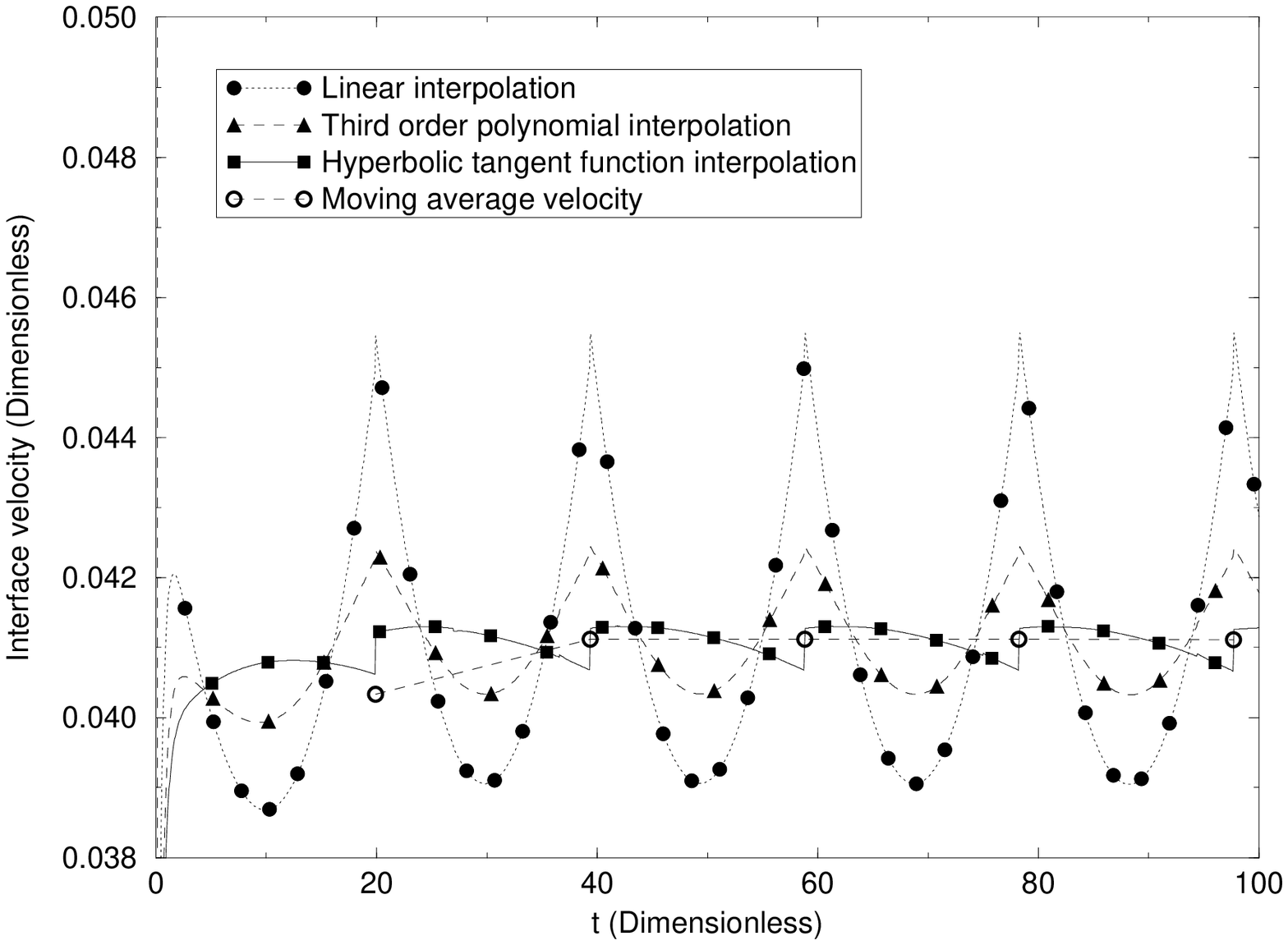}
\fi
\caption{Interface velocity versus time for 1-D test problem}
\label{TIPVEL1D}
\end{figure}


\ifDRAFT
 \clearpage
\fi
\begin{figure}[htb]
\ifDRAFT
 \centering
 \includegraphics[width=\colwidth]{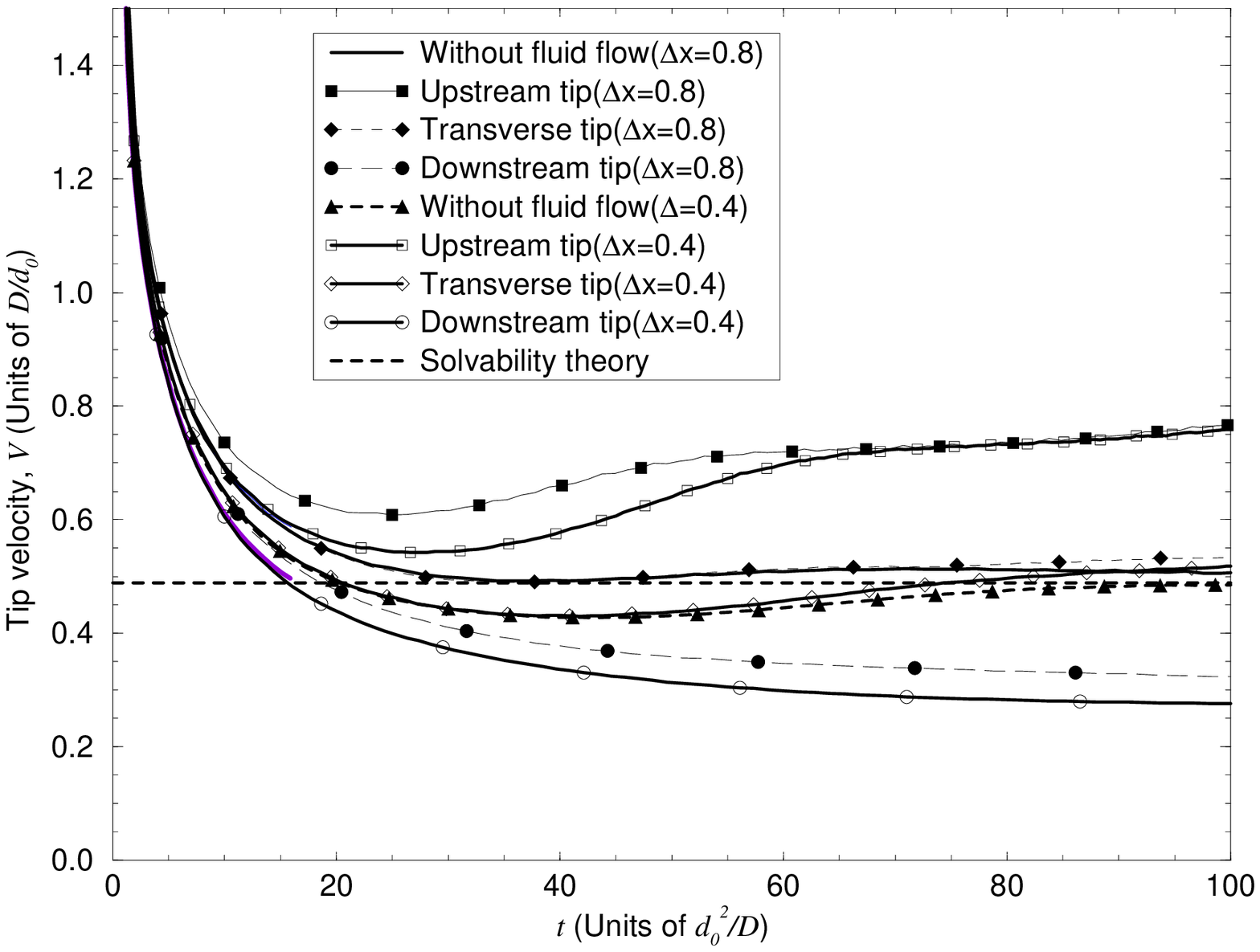}
 \includegraphics[width=\colwidth]{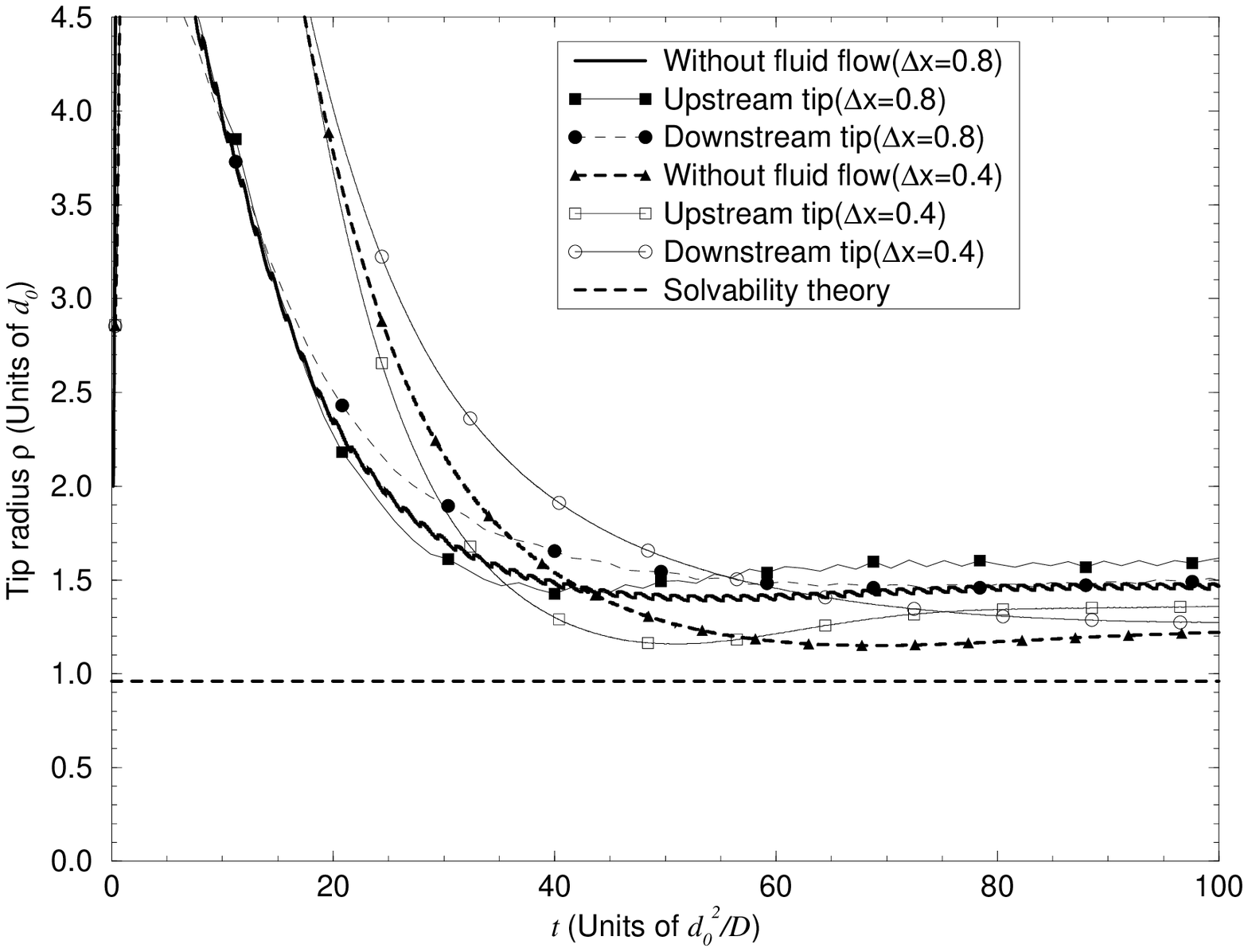}
\fi
\caption{Tip velocities and tip radii in 2D}
\label{tipvel2D}
\end{figure}

\ifDRAFT
 \clearpage
\fi
\begin{figure}[htb]
\ifDRAFT
 \centering
 \includegraphics[width=\colwidth]{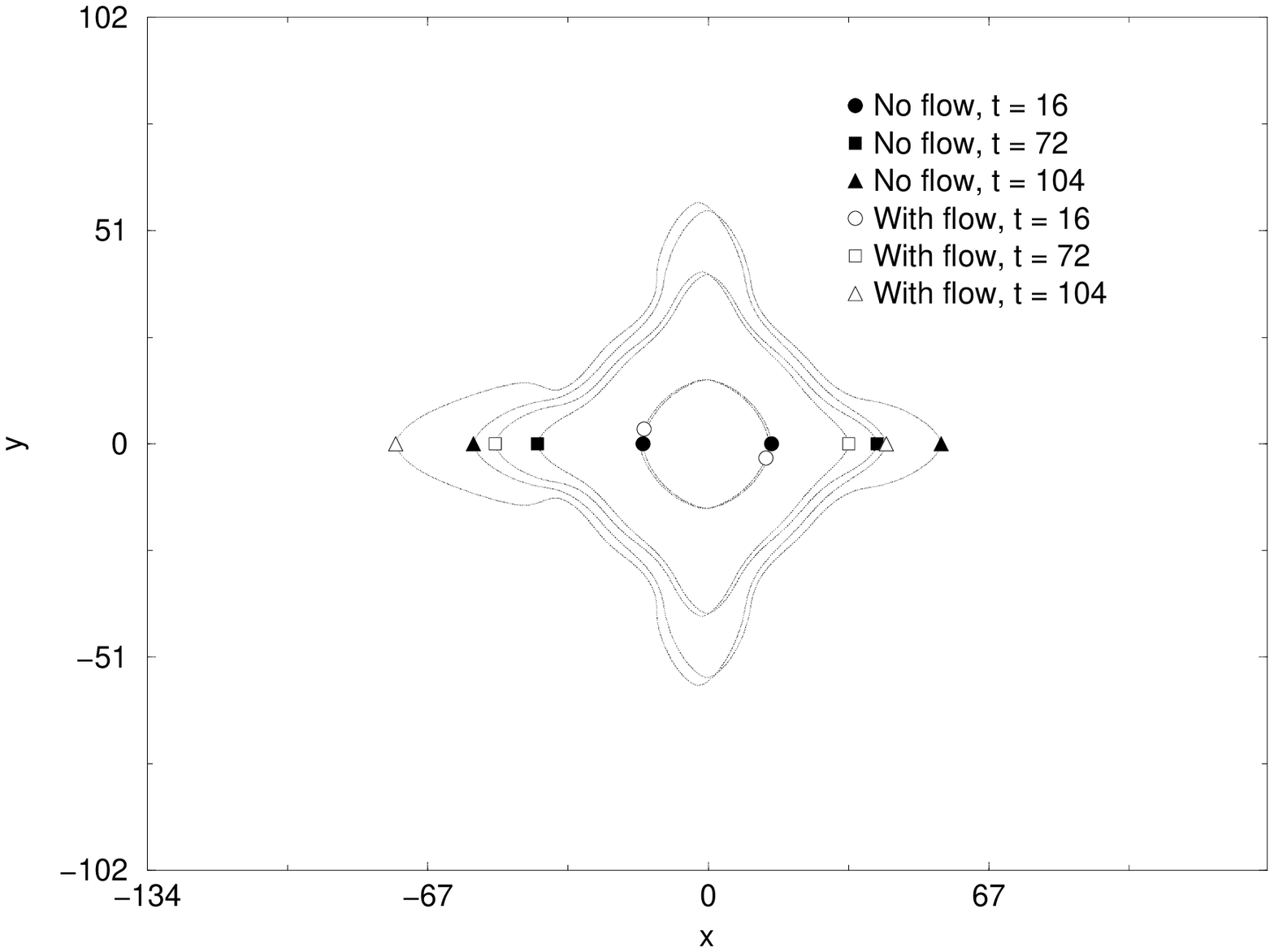}
\fi
\caption{Computed interfaces in 2D without and with fluid flow at
t=16, 72, and 104}
\label{2Dinter}
\end{figure}

\ifDRAFT
 \clearpage
\fi
\begin{figure}[htb]
\ifDRAFT
 \centering
 \includegraphics[width=\colwidth]{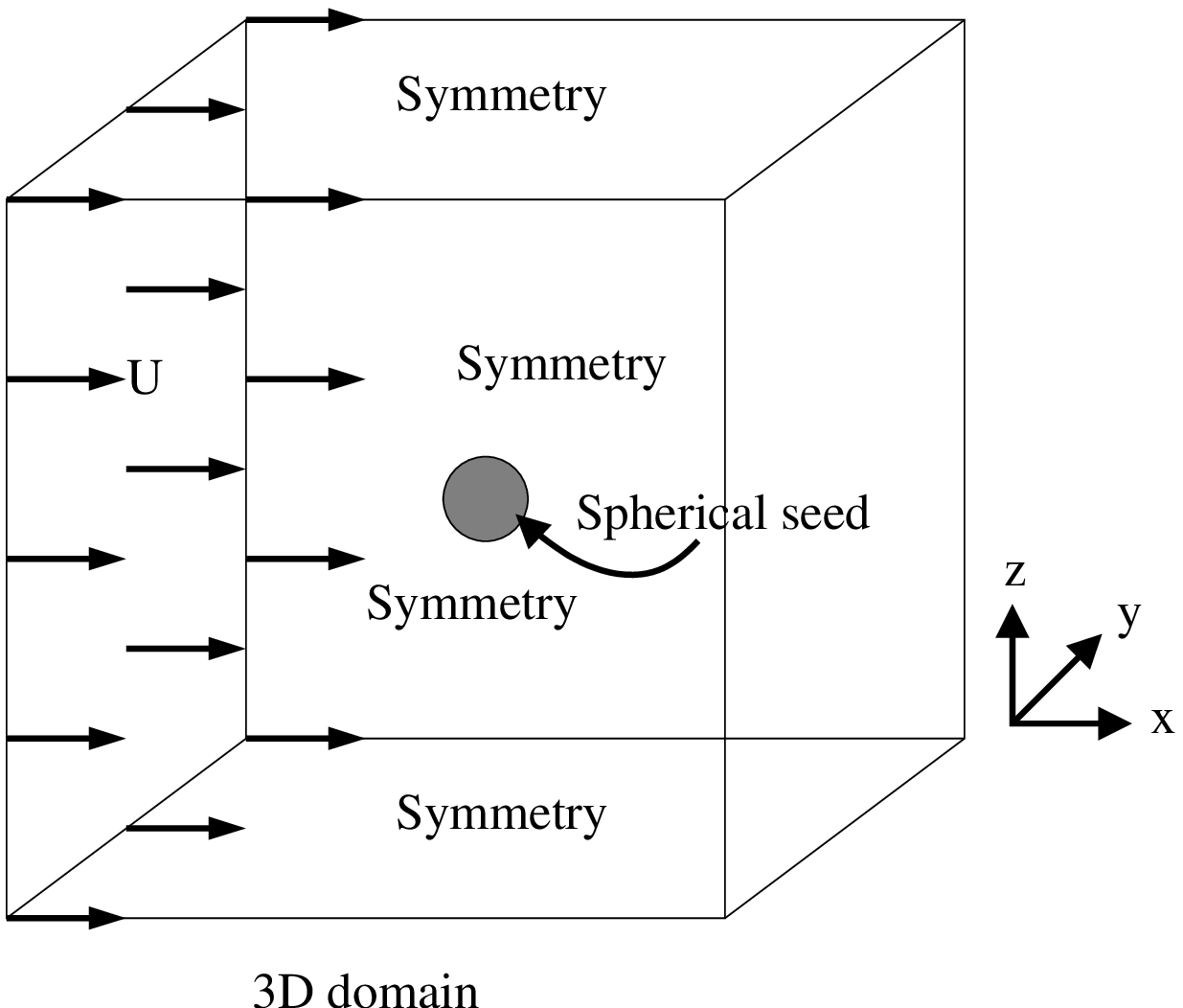}
\fi
\caption{Schematic diagram of 3D computation domain and boundary
conditions}
\label{3Ddomain}
\end{figure}

\ifDRAFT
 \clearpage
\fi
\begin{figure}[htb]
\ifDRAFT
 \centering
 \includegraphics[height=4in,angle=270]{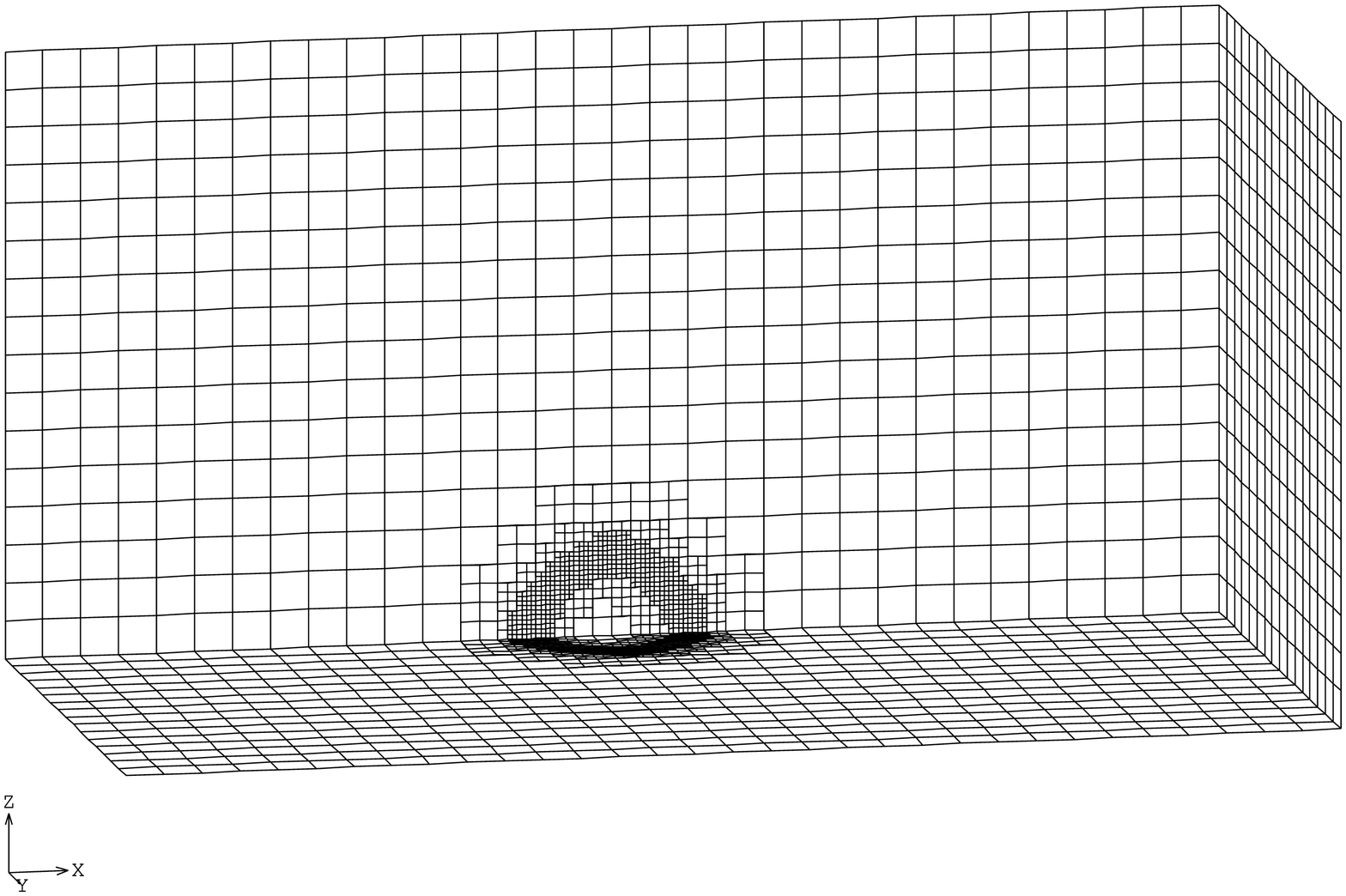}
 \includegraphics[height=4in,angle=270]{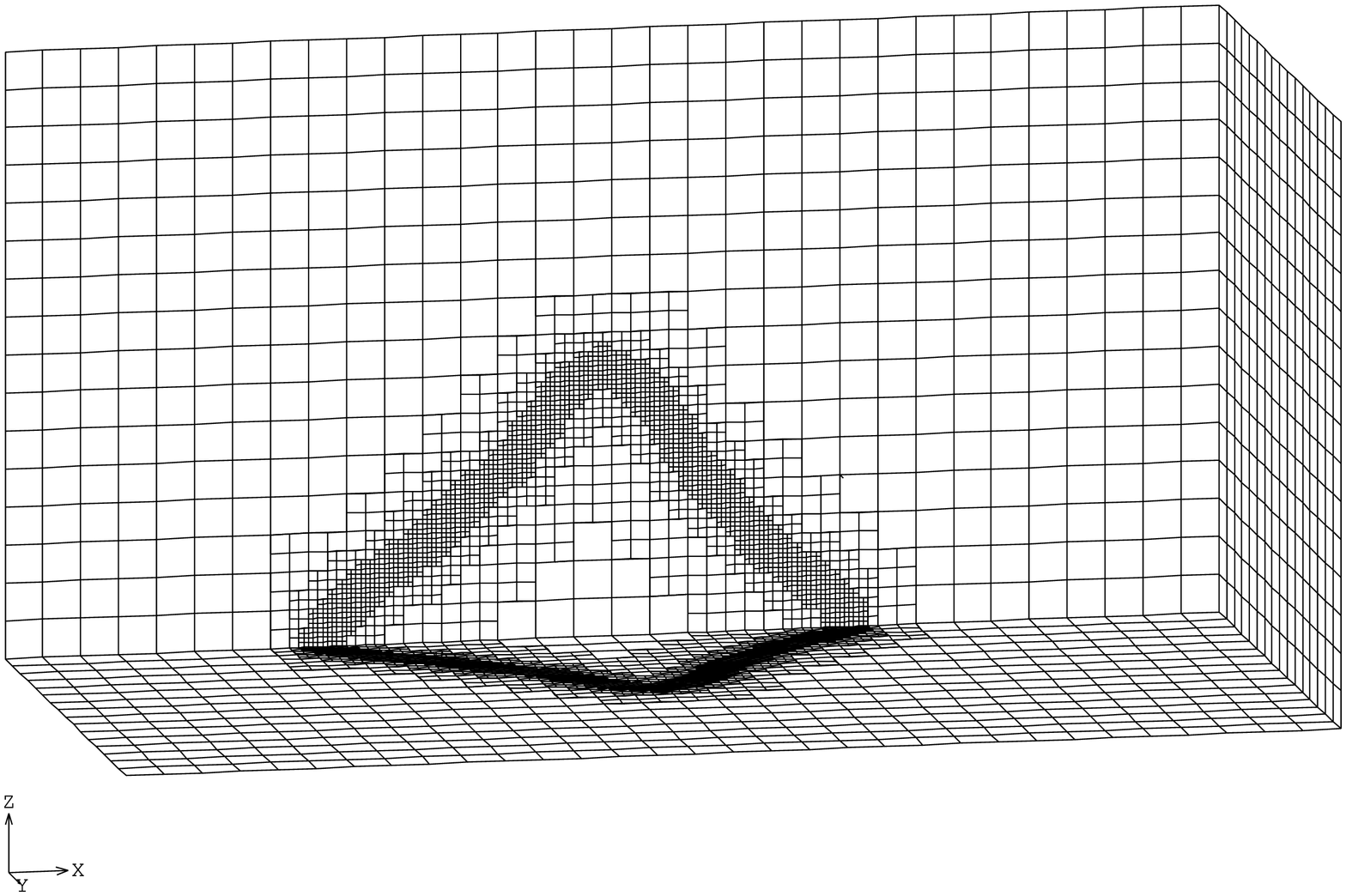}
 \includegraphics[height=4in,angle=270]{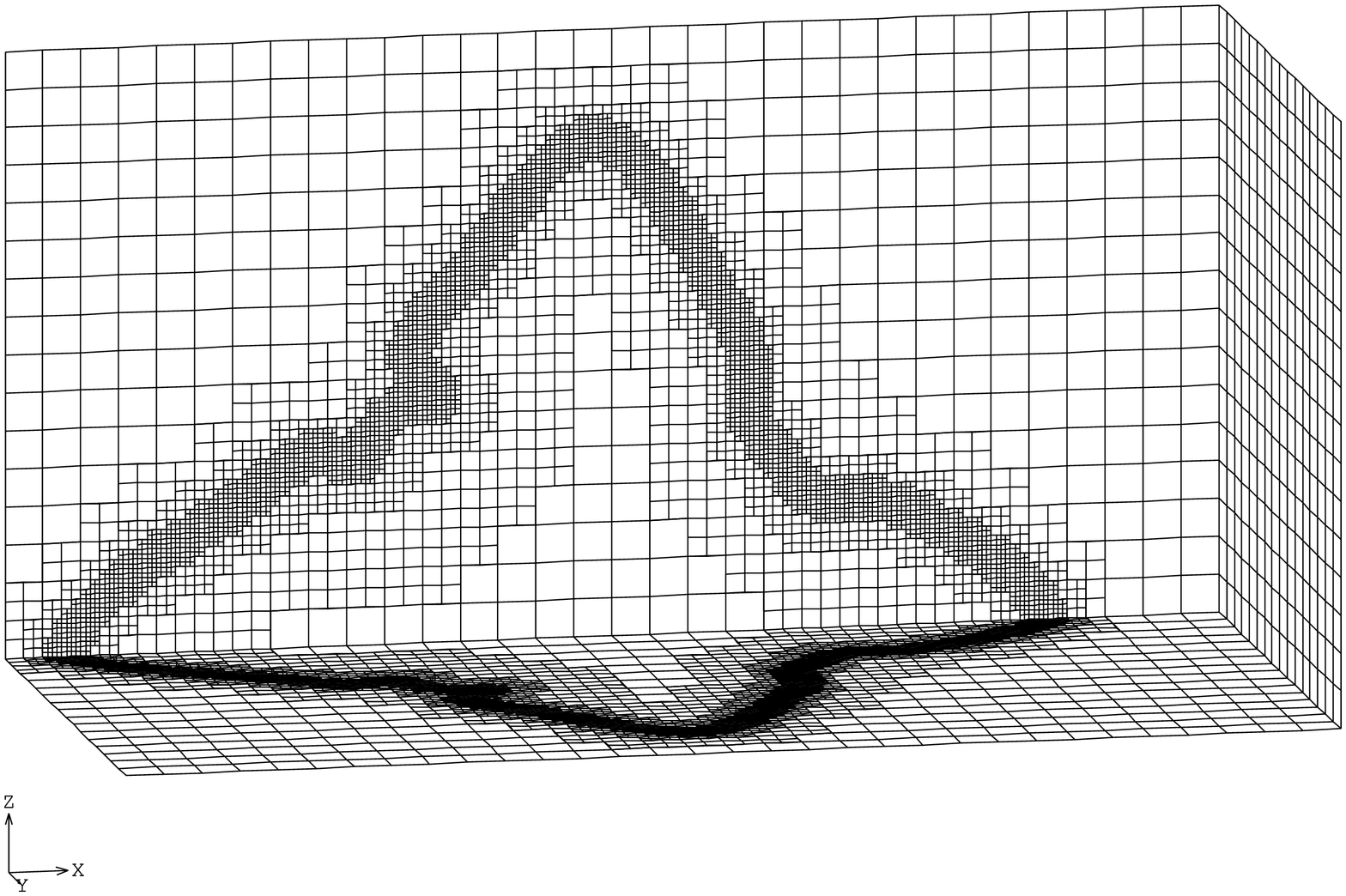}
\fi
\caption{Grid configurations of adapted grids t=9.6, 43.2, and 86.4}
\label{3Dgrid}
\end{figure}
%

%
\ifDRAFT
 \clearpage
\fi
\begin{figure}[htb]
\ifDRAFT
 \centering
 \includegraphics[width=\colwidth]{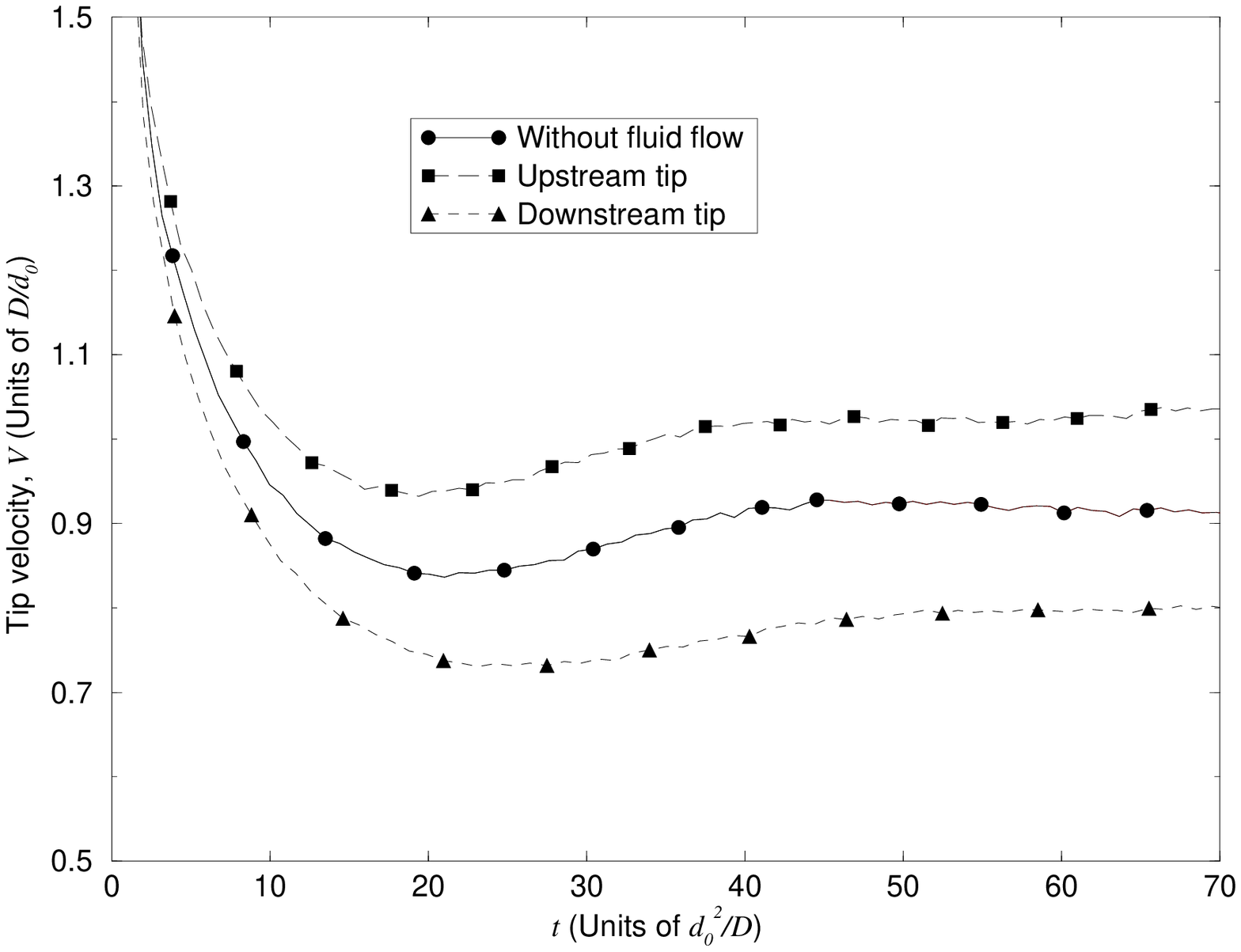}
 \includegraphics[width=\colwidth]{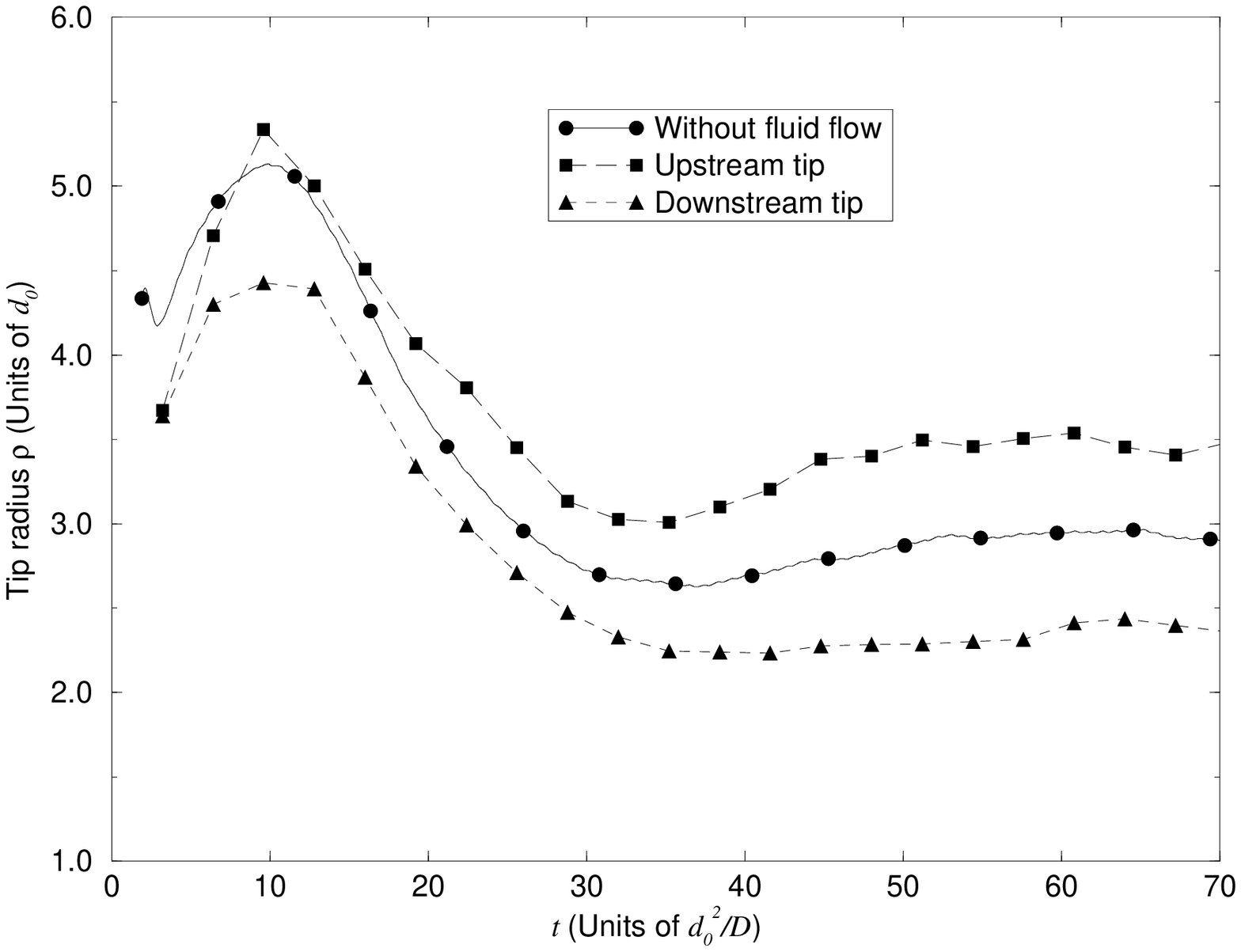}
\fi
\caption{Tip velocities and tip radii in 3D}
\label{tipvel3D}
\end{figure}
%
%
\ifDRAFT
 \clearpage
\fi
\begin{figure}[htb]
\ifDRAFT
 \centering
 \includegraphics[width=\colwidth,angle=0]{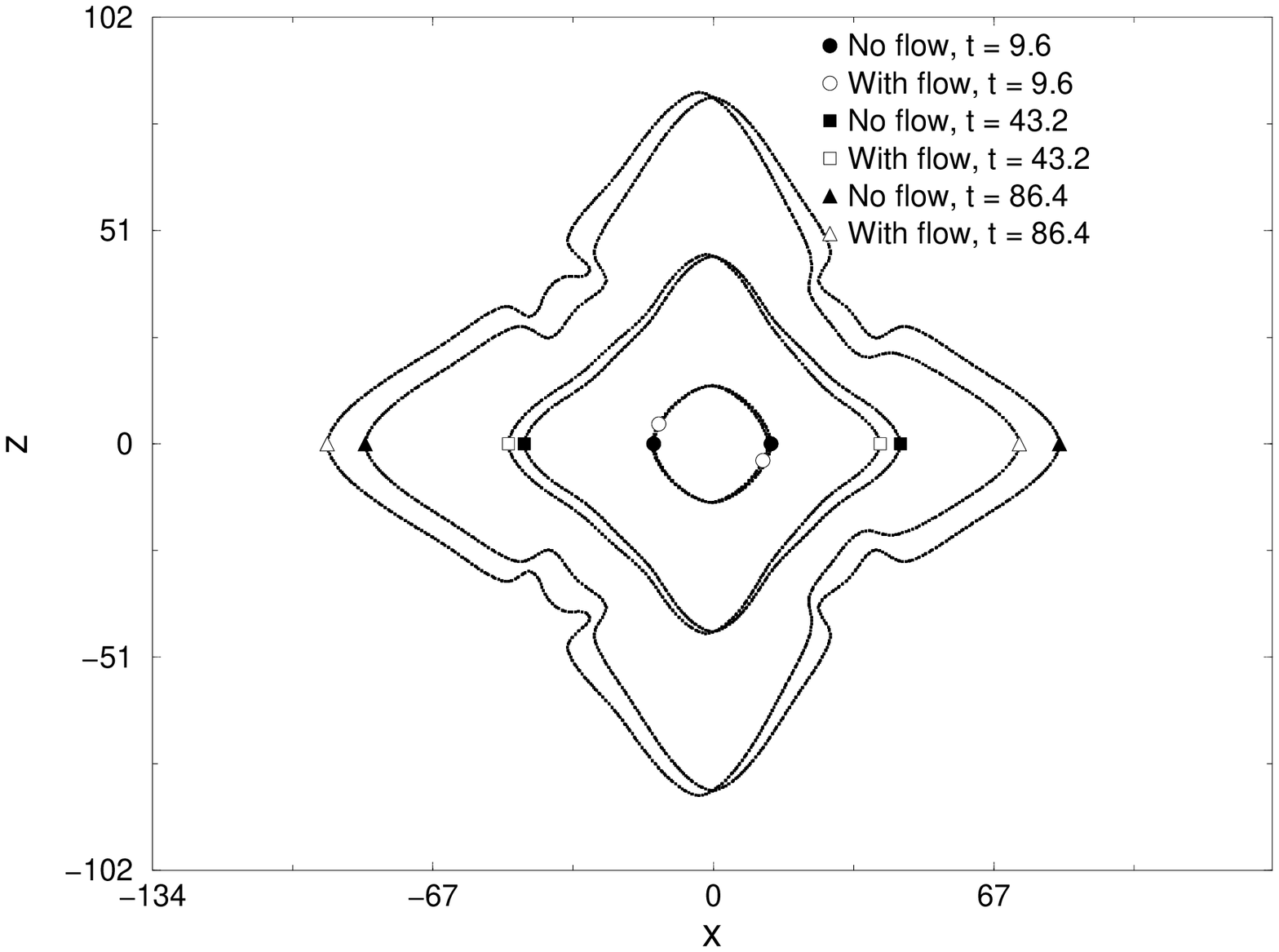}
\fi
\caption{Computed interfaces in 3D without and with fluid flow at
t=9.6, 43.2, and 86.4}
\label{3Dinter}
\end{figure}

\ifDRAFT
 \clearpage
\fi
\begin{figure}[htb]
\ifDRAFT
 \centering
 \includegraphics[width=\colwidth]{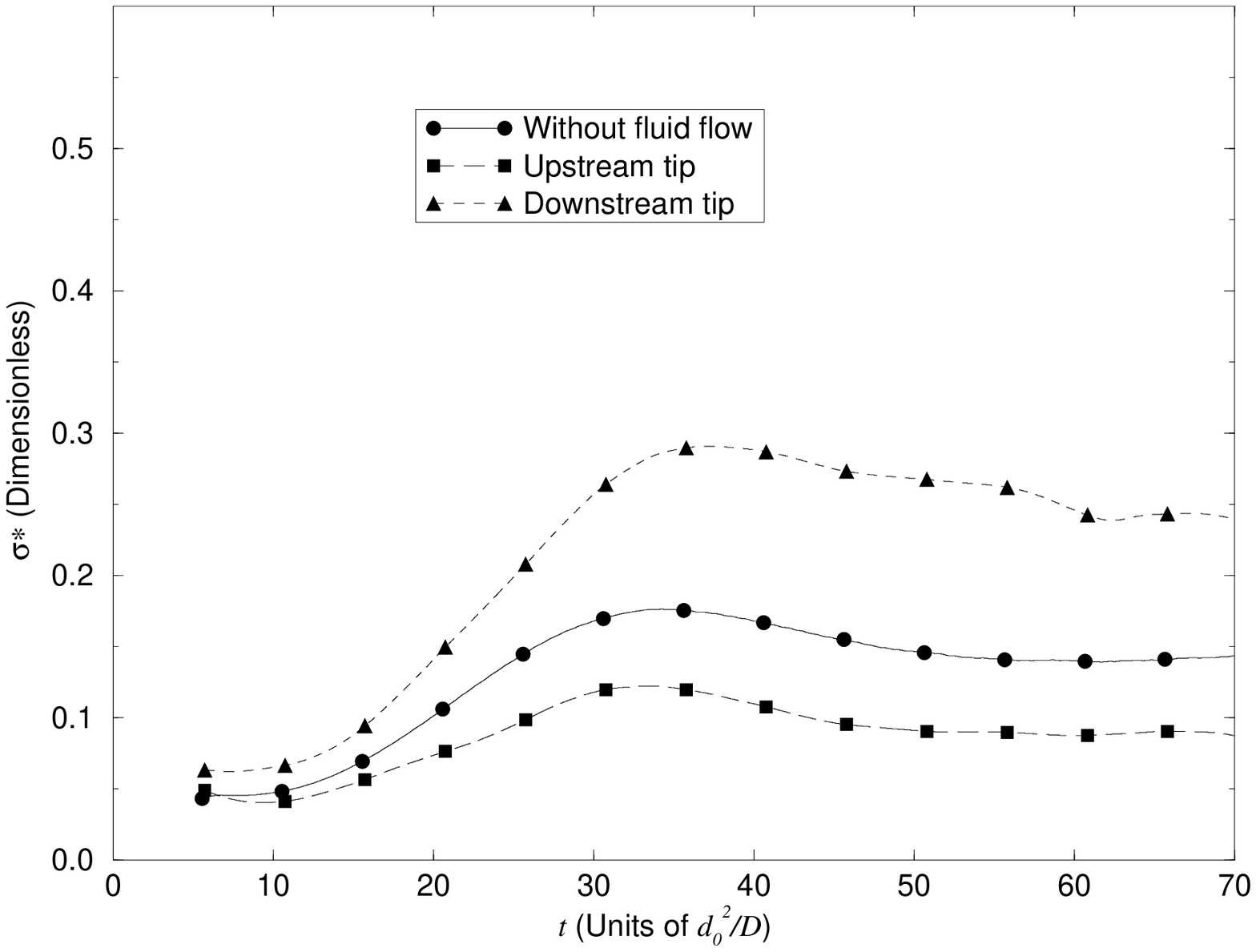}
\fi
\caption{$\sigma^*$s in 3D}
\label{sigma3D}
\end{figure}

\fi

\end{document}